\documentclass[aps,prb,twocolumn,floatfix,superscriptaddress,byrevtex,citeautoscript,10pt,amsmath,amssymb]{revtex4-2}

\usepackage{graphicx}
\usepackage{dcolumn}
\usepackage{bm}
\usepackage{xcolor}
\usepackage{textcomp}
\usepackage{booktabs}
\usepackage{soul}
\usepackage{import}
\usepackage{float}
\usepackage[pass]{geometry}

\newcommand{\Cu}{\mathrm{Cu}}
\newcommand{\N}{\mathrm{N}}

\newcommand{\defe}[2]{#1$_{\mathrm{#2}}$}
\newcommand{\defeq}[3]{#1$_{\mathrm{#2}}^{#3}$}

\begin{document}

\title{First-principles finite-size correction schemes for point defects of Cu$_3$N}

\author{Abdul M. Reyes}
\email{abreyesu@unal.edu.co}
\affiliation{Universidad Nacional de Colombia, Sede Tumaco, Tumaco, Nariño 764501, Colombia}
\affiliation{Departamento de Física, Grupo de Óptica Cuántica y Centro de Excelencia en Nuevos Materiales (CENM), Universidad del Valle, A.A. 25360, Santiago de Cali, Colombia}

\author{Sebastian E. Reyes-Lillo}
\email{sebastian.reyes@unab.cl}
\affiliation{Departamento de Física y Astronomía, Universidad Andres Bello, Santiago 837-0136, Chile}

\author{Eduardo Menéndez-Proupin} 
\email{emenendez@us.es}
\affiliation{
 Departamento de Física Aplicada I, Escuela Politécnica Superior, Universidad de Sevilla, Seville E-41011, Spain
}%

\date{\today}

\begin{abstract}
Point defects play a key role in determining semiconductor properties, such as electrical conductivity and photoluminescence, and often enable functional behavior. Accurate first-principles supercell simulations of point defects require reliable finite-size corrections. In this study, we combine PBE+U structural relaxations with HSE hybrid-functional calculations to determine defect formation energies and thermodynamic transition levels of Cu$_3$N. Finite-size trends are quantified using $\Gamma$-point calculations in supercells containing 256, 864, and 2048 atoms. We assess and extend the Makov-Payne and Lany-Zunger correction schemes by introducing additional $1/L^n$ terms, together with core-level potential alignment and defect-specific scaling models. Using Cu$_3$N as a case study, we show that charged vacancies are accurately described by the Makov-Payne-type scaling ($1/L + 1/L^{3}$), whereas interstitial defects with shallow or weakly localized electronic character are better captured by a hydrogenic impurity model that accounts for defect-band dispersion. Residual trends for neutral or weakly localized defects are described by higher-order polynomial fits in $1/L^{3}$ and $1/L^{4}$. Hybrid-functional energetics corrected using PBE+U-based finite-size trends confirm the intrinsic $p-n$ doping character of Cu$_3$N under the conditions considered and demonstrate that no single finite-size correction can be transferred across all defect types.
\end{abstract}

\maketitle


\section{Introduction\label{sec:intro}}

Point defects in semiconductors correspond to single lattice site imperfections in an otherwise periodic crystal. These include interstitials (X$_i$) and vacancies (V$_\mathrm{X}$) of an atom $X$, as well as substitutions (Y$_\mathrm{X}$) of a host atom $Y$ by atom $X$. Intrinsic defects comprise cases in which the species involved are the same as those that form the host material, while extrinsic defects include species not present in the pristine crystal. Typically, defect concentrations are low, e.g., parts per million, and are driven by Maxwell-Boltzmann thermodynamics. Despite their low concentration, point defects are responsible for key properties of semiconductors, such as electrical conductivity and photoluminescence spectra~\cite{Zheng2021, Zhang2021}, which, in turn, have important implications and functional properties in modern electronics~\cite{Seebauer2010}, solar energy conversion~\cite{Ran2020, Wang2018b, Bai2018}, and batteries~\cite{Zhang2020a}.

Theoretical studies of point defects generally rely on first-principles density functional theory (DFT)-based calculations and the supercell approach~\cite{Freysoldt2014}, with varying degrees of sophistication~\cite{Alkauskas2011}. Within the DFT supercell approach, practical cell sizes correspond to defect concentrations that are far above the dilute limit, leading to artificial interactions between the defect and its periodic images. Consequently, the calculated formation energies of charged defects are affected by (i) spurious long-range Coulomb interactions and (ii) the arbitrary choice of the average electrostatic potential, which shifts the energy scales of supercells with different charges and/or numbers of atoms. To recover the dilute limit, we therefore introduce a defect-dependent finite-size correction $\Delta E$ to the formation energy.

\begin{figure}[tbh!]
\includegraphics[width=\columnwidth]{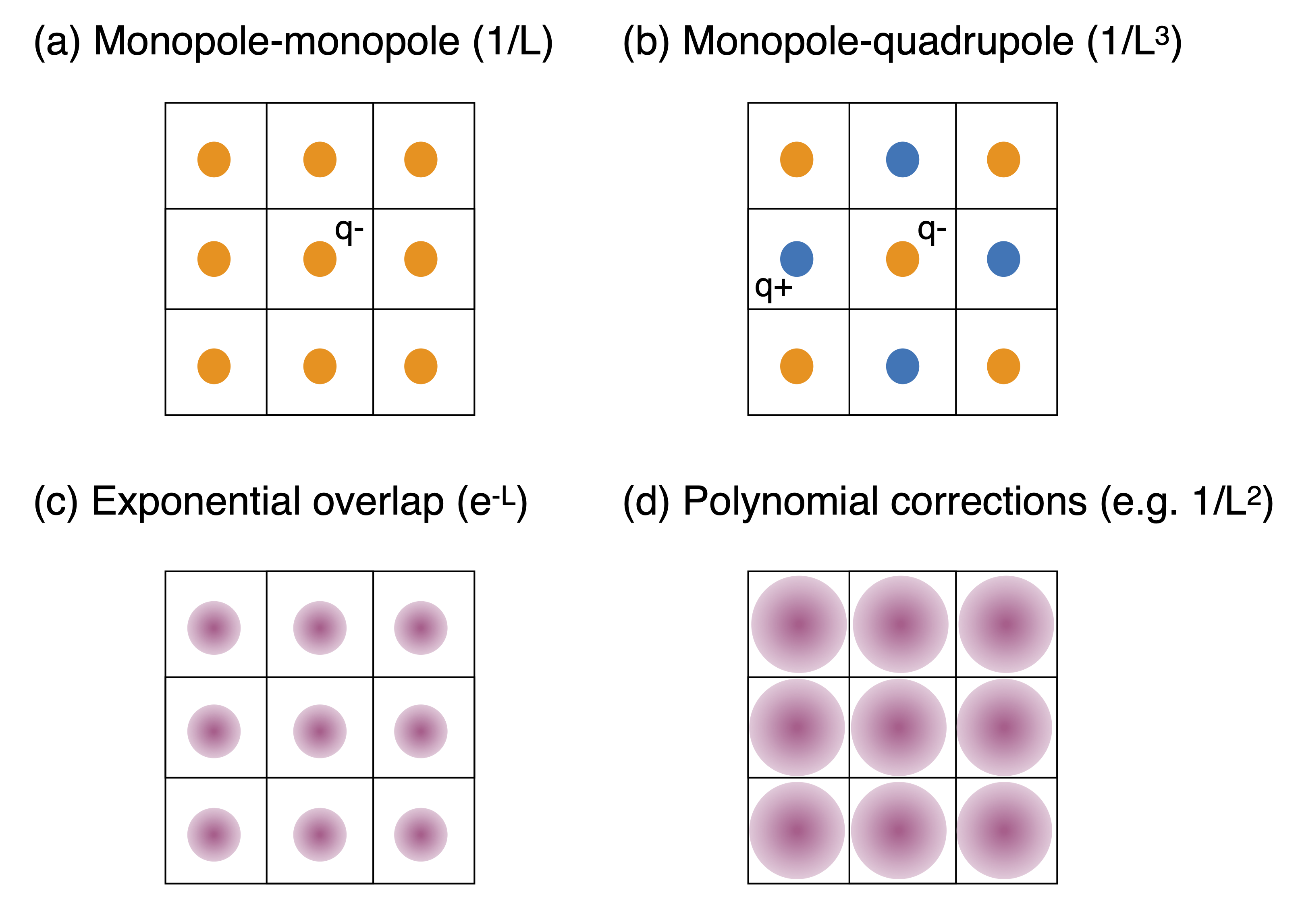}
\caption{
Dominant finite-size interactions between a defect and its periodic images: (a) monopole-monopole ($\propto 1/L$) and (b) monopole-quadrupole ($\propto 1/L^3$) terms for localized charged defects, (c) exponential defect-wavefunction term ($\propto e^{-L}$) for defects with overlapping localized electronic states, and (d) higher-order polynomial corrections for weakly localized or neutral defects.}
\label{models1}
\end{figure}

Figure~\ref{models1} shows dominant spurious electrostatic interactions between a charged defect and its periodic images~\cite{Lambrecht2011, Walsh2021}. The seminal work of Makov and Payne~\cite{Makov1995} assumes a strong excess charge around the defect and incorporates quadrupole effects (see Fig.~\ref{models1}(a) and (b)). The finite-size correction of Makov-Payne is obtained from a multipole expansion, retaining the monopole–monopole ($\propto 1/L$) and monopole–quadrupole ($\propto 1/L^3$) terms, as
\begin{equation}
\label{eq:MP}
\Delta E_{M-P}=\frac{\alpha_M q^2}{2 \varepsilon L}+ \frac{2 \pi q Q }{3 \varepsilon L^3},
\end{equation}
where $q$ is the defect charge, $\alpha_M$ is the lattice Madelung constant of the supercell, $\varepsilon$ is the (static) dielectric constant of the host, $L \equiv V^{1/3}$ is the characteristic linear size of the supercell, and $Q$ is the quadrupole moment of the defect charge distribution. Several alternative finite-size correction schemes have also been proposed~\cite{Dabo2008,Freysoldt2009,Lany2008EMP, Kumagai2014}. In addition, the relative zero of the formation energy, originating from the mismatch of charge and stoichiometry among different supercell calculations, is corrected by adding the ``potential alignment'' term
\begin{equation}
\label{eq:PA}
\Delta E_{PA}= q (V - V_0),
\end{equation}
where $q$ is the defect charge, and $V$ is the far-field electrostatic potential offset of the defective cell with respect to the pristine reference $V_0$, commonly evaluated using core-level–averaged potentials~\cite{Lany2008,Freysoldt2009}.

Notably, test calculations with very large supercells show that, for a number of defects,
the combination of the Makov-Payne correction and the potential alignment converges towards the dilute limit~\cite{Lany2008EMP}.  These results also show that the monopole-quadrupole
correction can be proportional to the monopole-monopole correction, implying a dependence of the defect quadrupole  $Q\sim -qL^2$. Therefore, the finite-size correction method of Lany and Zunger proposes 
\begin{equation}
\label{eq:LZ}
\Delta E_{L-Z}=\Delta E_{PA} + (1+f)\frac{\alpha_M q^2}{2 \varepsilon L},
\end{equation}
where the potential alignment correction $\Delta E_{PA}$ is typically proportional to $1/L^3$, and the factor $f\simeq -1/3$ accounts for the proportionality between the monopole-quadrupole and monopole-monopole terms. 

Another popular approach is the finite-size (FS) scaling method~\cite{Lany2008EMP, Castleton2004, Castleton2009, Komsa2012PRB, Komsa2012Physica}, which involves calculating the defect finite-size correction energy by employing a sequence of supercells with increasing size $L$ and fitting the defect formation energy with the function 
\begin{equation}
\label{eq:HFS}
H(L) = H(\infty) + \frac{\gamma_1}{L} + \frac{\gamma_3}{L^3}.
\end{equation} 
Eq.~\eqref{eq:HFS} implies a size correction
\begin{equation}
\label{eq:FS}
\Delta E_{FS} (L) = H(\infty)-H(L)=  -\frac{\gamma_1}{L} - \frac{\gamma_3}{L^3},
\end{equation}
where $\gamma_1$ and $\gamma_3$ are obtained as fitting parameters. Let us note that if Eqs.~\eqref{eq:MP} or \eqref{eq:LZ} apply, then $\gamma_1<0$. 

While the methods of Makov-Payne and Lany-Zunger (see Eqs.~\ref{eq:MP} and \ref{eq:LZ}) have the advantage of offering a fast correction method without the need for a sequence of computationally expensive calculations with larger supercells, the finite-size scaling method (see Eq.~\eqref{eq:FS}) allows for the assessment of size effects for defects that have not been explored previously. 

Previous DFT studies using the supercell approach and the correction methods described above have critically examined the performance and applicability of existing \textit{a posteriori} schemes, including the Makov-Payne, Lany-Zunger, and Freysoldt-Neugebauer-Van de Walle methods~\cite{Freysoldt2009}, and have established the conditions under which these approaches can be reliably applied, concluding that the scheme of Freysoldt-Neugebauer-Van de Walle generally provides the best performance for defects with localized electronic states, regardless of the material~\mbox{\cite{Komsa2012PRB, Komsa2012Physica}}. This body of work has enabled accurate calculations of defect properties relevant to photovoltaics, photocatalysis, and batteries~\cite{Kokott2018,Kumagai2014,Durrant2018}. 

However, despite this progress, several classes of point defects still exhibit deviations from these standard models, and their formation-energy corrections are not well described by Eqs.~\ref{eq:MP}--\ref{eq:FS}~\cite{Walsh2021}. These deviations can arise from additional long-range interactions not captured by the Makov-Payne and Lany-Zunger schemes, such as monopole-dipole interactions (\mbox{$\propto 1/L^2$}), dipole-dipole interactions in the neutral case (\mbox{$\propto 1/L^3$}), and residual interactions (\mbox{$\propto 1/L^4$}). In addition, short-range interactions originating from lattice stress (\mbox{$\propto 1/L^3$}) and defect-wavefunction overlap (\mbox{$\propto e^{-L/L_0}$}) may not be sufficiently attenuated in commonly used small to medium supercells and may also require further correction.

Importantly, even neutral point defects may retain sizable formation-energy errors in medium to large-sized supercells~\cite{Castleton2004}. Elastic interactions and wave-function orthogonality constraints between periodic defect images may produce abnormally slow convergence of formation energies with supercell size~\cite{Freysoldt2022, Corsetti2011}, requiring additional $1/L$ polynomial terms to correct for these effects~\cite{Castleton2009}. In some cases, such polynomial corrections prove insufficient, and explicit tests with larger supercells are still required~\cite{Burr2017}. Several other works highlight the relevance of finite-size scaling corrections in specific contexts~\cite{Gallino2010, Erhart2006, Rurali2010}.
 
In this work, we use first-principles methods to investigate the performance of the Makov-Payne and Lany-Zunger finite-size correction methods when additional $1/L^n$ terms are included. As a case study, we investigate the low-energy defect properties of copper nitride (Cu$_3$N), a material of interest for solar energy conversion and optoelectronic devices~\cite{Jiang2018,Lindahl2018,Lu2013,Yang2014}. Cu$_3$N can host a diverse array of defect types and, therefore, provides a useful platform for identifying distinct finite-size behaviors among different classes of defects within a single host. From this analysis, we extract practical considerations for selecting an appropriate correction scheme when computational limitations prevent the use of sufficiently large supercells.

We generalized the previous discussion by considering defects with overlapping and weakly localized electronic states (Figs.~\ref{models1}(c) and~\ref{models1}(d) resp.). Defects with weakly localized electronic states, such as shallow defects, present inherent methodological challenges due to insufficient supercell size or the use of \textbf{k}-point grids. In the case of highly dispersive bands, band-filling corrections can be used to compensate for size-induced occupation artifacts~\cite{Lany2008EMP}. Although these corrections are relatively simple to apply, size effects and band-filling corrections may also lead to incorrect minimal energy structure geometries~\cite{Menendez-Proupin2015}. Overlapping localized (e.g. deep in-gap) defects whose impurity bands exhibit a pronounced \textbf{k}-dependent dispersion tend to deviate from simple polynomial finite-size fits. As will be discussed in more detail below, in such cases, the artificial size effects mainly arise from the widening of the impurity band in relatively small supercells.

For the case of weakly localized electronic states, which show a mild dispersive in-gap band, we propose the polynomial fitting model (PFM) given by
\begin{equation}
H_{PFM}(L) = H(L_0)+B_0 + \sum_{n=1}^{4} \frac{B_n}{L^n},
\label{eq:pfm}
\end{equation}
where $L_0$ is a convenient supercell size that will be discussed in Sec.~\ref{Sec:DefectDFT}.
Compared to Eq.~\eqref{eq:HFS}, Eq.~\eqref{eq:pfm} introduces additional powers of $1/L$.  
Eq.~\eqref{eq:pfm} implies that the size correction to $H(L_0)$ is the fitting parameter $B_0$. For arbitrary $L$, the size correction is 
\begin{equation}
\Delta E_{PFM} (L) =H_{PFM}(\infty)-H_{PFM}(L)= -\sum_{n=1}^{4} \frac{B_n}{L^n}.
\label{models}
\end{equation}

If the condition  $H_{PFM}(L_0)=H(L_0)$ is enforced, then
\begin{equation}
\label{eq:pfmequiv}
\Delta E_{PFM} (L_0) =B_0=-\sum_{n=1}^{4} \frac{B_n}{L^n_0}.
\end{equation}

Here, due to computational limitations, only three values of $1/L$ are available. Therefore, only $B_0$ and two additional $B_n$ coefficients are allowed to be nonzero in the fitting, while Eq.~\eqref{eq:pfmequiv} remains satisfied. For neutral and shallow defects, where the induced charge is delocalized, $B_1 = 0$. Moreover, the equivalence between Eqs.~\eqref{eq:MP} and \eqref{eq:LZ} implies that if Eqs.~\eqref{eq:MP} or \eqref{eq:LZ} apply, then $B_1<0$. These constraints allow us to discard some fitting models.

For defects that generate a dispersive in-gap band, the finite-size correction is modeled as
\begin{equation}
\label{eq:expo}
\Delta E_{EXP-L}(L) = A\left(1+\frac{L}{B}\right)e^{-L/B},
\end{equation}
where $A$ and $B$ are fitting parameters obtained by fitting the formation energies as a function of the supercell size $L$. This functional form is motivated by a tight-binding description of a defect-derived impurity band on a cubic lattice~\cite[Ch.~9]{Kittel2005}, whose dispersion is given by
\begin{equation}
 \epsilon(\mathbf{k}) = -\epsilon_0 - 2\gamma(L) \sum_i \cos(k_i a),   
\end{equation}
where the sum runs over $i = x,y,z$, $\epsilon_0$ is the on-site self-energy, and $\gamma(L)$ is a size-dependent hopping parameter that can be obtained from a hydrogenic impurity wave function~\cite[Ch.~9, Eq.~11]{Kittel2005}. Because standard calculations sample the artificial dispersion of the defect superlattice at the supercell center, the impurity-band contribution to the total energy at the supercell $\Gamma$-point becomes $\epsilon(\mathbf{0}) = -\epsilon_0 - 6\gamma(L)$. Therefore, the correction $\Delta E_{EXP-L}(L)$ is approximately equal to $6\gamma(L)$ multiplied by the number of electrons occupying the defect-derived band at $\Gamma$.

\begin{figure}[tbh!]
\includegraphics[width=0.9\columnwidth]{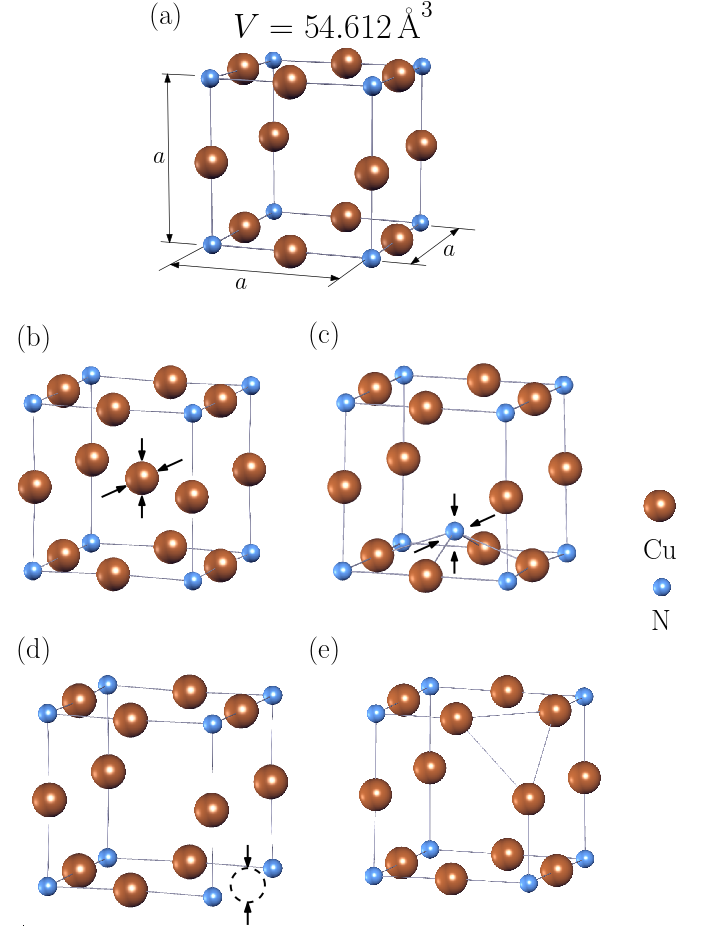}
\caption{(a) Crystal structure of cubic Cu$_3$N with space group $Pm\bar{3}m$ and relaxed lattice parameter of 3.794~{\AA}. Cu and N ions are denoted as gold and blue spheres. Lowest energy structures for (b) Cu$_{\textrm{i}}$, (c) N$_{\textrm{i}}$, (d) V$_{\textrm{Cu}}$ and (e) V$_{\textrm{N}}$. In (c), N$_{\textrm{i}}$ is located near the face center of the unit cell and exhibits a slight displacement towards the center of the cubic cavity. In (b), Cu$_{\textrm{i}}$ is located near the center of the unit cell, with minimal lattice distortion.}
\label{structures}
\end{figure}

\section{Application to $\mathrm{Cu}_{3}\mathrm{N}$}

$\mathrm{Cu}_{3}\mathrm{N}$ is a narrow gap semiconductor with multiple functional properties in bulk and thin film form, including solar energy conversion~\cite{Rodriguez-Tapiador2023, Rodriguez-Tapiador2023a}, optical information storage, microelectronic semiconductors~\cite{Jiang2018}, and electronic devices~\cite{Fioretti2016, Zervos2020}. Cu$_3$N has a cubic $Pm\bar{3}m$ structure at room temperature (see Fig.~\ref{structures}(a)) and exhibits small direct and indirect optical gaps of 1.4~eV and 1.0~eV, respectively~\cite{Zakutayev2014, Fioretti2016, Zervos2020}, high charge carrier mobilities (up to 200 cm$^2$/Vs), and a high optical absorption coefficient of 10$^5$ cm$^{-1}$ above 2 eV~\cite{Matsuzaki2018}. Cu$_3$N thin films display bipolar $n$-type and $p$-type doping characteristics as a function of growth temperature and Cu/N flux ratio, enabling the fabrication of $p$–$n$ junctions~\cite{Zakutayev2014, Matsuzaki2014}. The electronic structure of Cu$_3$N exhibits anti-bonding and bonding characteristics at the valence and conduction edges, respectively, suggesting defect tolerance~\cite{Zakutayev2014}. The large cavity of the body-centered cubic structure allows extrinsic doping with Ni~\cite{Lindahl2018}, Ag, Au~\cite{Lu2013}, Pd, In, Zn~\cite{Ji2010}, Fe, Co, Ni~\cite{Yang2014}, O~\cite{Zervos2021}, and Li~\cite{Wang2017a}. 

Previous DFT work has investigated the defect properties of $\mathrm{Cu}_{3}\mathrm{N}$~\cite{Fioretti2016, Caskey2014, Matsuzaki2018, Yee2018, Madhu2026}. Despite being thermodynamically metastable against decomposition in bulk copper and molecular nitrogen~\cite{Caskey2014}, Cu$_3$N has been predicted to be dynamically stable~\cite{Yu2005}. The conduction band minimum (CBM) and the valence band maximum (VBM) of $\mathrm{Cu}_{3}\mathrm{N}$ are located at the M ($\mathbf{k} = (\pi/a)(1,1,0)$) and R ($\mathbf{k} = (\pi/a)(1,1,1)$) points, respectively, of the Brillouin zone~\cite{Moreno-Armenta2007, Yee2018}. Cu$_{\textrm{i}}$ and V$_{\textrm{Cu}}$ display defect levels near the CBM and VBM, respectively~\cite{Peng2013, Yee2018}, supporting the observation of $p-n$ semiconductor transition. Antisite defects such as Cu$_{\textrm{N}}$ and N$_{\textrm{Cu}}$ exhibit high formation energies due to the significant electronegativity difference between Cu and N. The large Coulombic repulsion between the defects and the lattice renders the antisite defects inherently unstable~\cite{Yee2018, Zervos2020, Zervos2021}. 

Here, we focus on the formation energies of the thermodynamically stable intrinsic defects, evaluated under several supercell–size correction schemes. To that end, semilocal and hybrid density-functional calculations are combined with three complementary finite-size correction schemes to elucidate the $p$–$n$ doping behavior of Cu$_3$N. The corrected formation energies and charge-transition levels  of intrinsic defects are evaluated under both N-rich and Cu-rich chemical potentials. Under N-rich growth conditions, charge neutrality is governed by \defe{V}{Cu}, \defe{Cu}{i}, and \defe{N}{i}, yielding \(p\)-type conductivity, whereas in Cu-rich conditions, \defe{Cu}{i} becomes the most stable defect and drives \(n\)-type conductivity. Because the spatial extent of these defects varies widely, no single size-correction approach is applicable to all cases. The finite-size correction methods developed here provide guidelines for choosing correction strategies when defect wave functions extend over multiple length scales, yielding dilute-limit energetics without excessive computational cost.

\section{Computational details}

\subsection{Electronic reference energies}
Spin-polarized DFT calculations were performed using the Vienna Ab-initio Simulation Package (VASP)~\cite{Kresse1996, Kresse1996c}. We used the generalized gradient approximation of Perdew, Burke, and Ernzerhof (PBE)~\cite{Perdew1996}, supplemented by the Hubbard $U$ method~\cite{Dudarev1998} (PBE+U), to describe the Cu $d$ orbitals more accurately. The Hubbard U value for the Cu $3d$ states was set to 5~eV~\cite{Fioretti2016}, using a plane-wave kinetic-energy cutoff of 400~eV and performing calculations at the $\Gamma$-point. Core–valence interactions were treated with projector-augmented wave potentials that included 11 and 5 valence electrons for Cu and N, respectively~\cite{Joubert1999}. Structures were relaxed within PBE+U until the maximum residual force was below 0.01~eV/\AA. For cubic Cu$_3$N, the optimized lattice constant is $a=3.794$~\AA, in close agreement with the experimental value of 3.815~\AA\ (deviation of $\sim$0.6\%)~\cite{Zervos2020}. 

Previous work has shown that semilocal functionals underestimate the optical band gap of Cu$_3$N by $\sim40\%$~\cite{Moreno-Armenta2007, Navio2007, Peng2013}. By contrast, higher levels of theory, such as hybrid functionals~\cite{Peng2013,Yee2018} and the GW approximation within many-body perturbation theory~\cite{Peng2013,Zakutayev2014}, are able to reproduce the experimental optical gap~\cite{Ebaid2020, Zakutayev2014}. Consequently, defect formation energies were obtained from static total-energy calculations using the Heyd–Scuseria–Ernzerhof (HSE) hybrid functional~\cite{Heyd2003,Krukau2006}. Our HSE band structure yields an indirect band gap of 0.93~eV (see Fig.~S1 in Supplemental Material (SM)~\cite{SI_Cu3N}). For the $6\times6\times6$ supercell, HSE (PBE+U) gives a gap of 0.92~eV (0.67~eV) when extracted from Kohn-Sham eigenvalues and 0.81~eV (0.57~eV) when extracted from total-energy differences between the neutral and $+1/-1$ charge states. These values are consistent with previous \textit{ab initio} and experimental results~\cite{Moreno-Armenta2007, Peng2013}.

\begin{figure}[tbh!]
    \centering
    \includegraphics[width=0.8\linewidth]{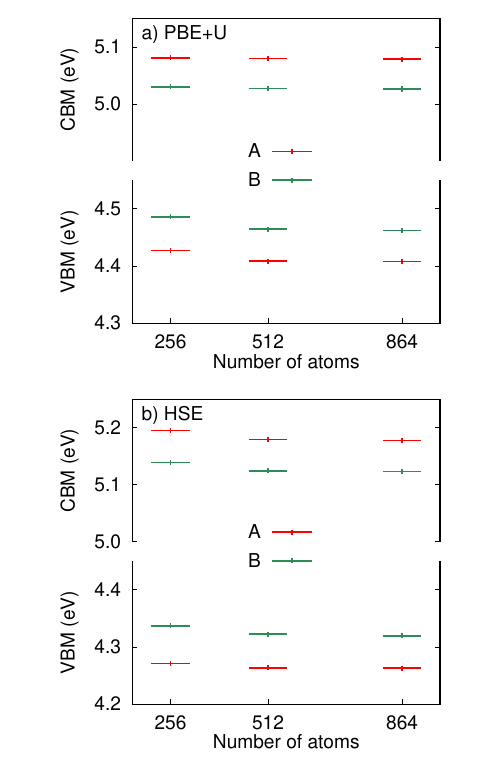}
    \caption{Comparison of two methods for determining the band edges and band gap, in (a) using PBE+U and (b) HSE functionals. In method~A, the CBM and VBM are obtained from the generalized Kohn–Sham eigenvalues of the neutral pristine supercell. In method~B, the CBM and VBM are extracted from total-energy differences between neutral and charged supercells with $q=-1$ and $q=+1$, respectively. A band-alignment correction is applied in method~B, but its magnitude is always $\leq 0.01$~eV.}
    \label{fig:sizeeffectgap}
\end{figure}

Previous work has shown that generalized Kohn–Sham eigenvalues obtained with HSE hybrid functionals can suffer from finite-size effects, leading to inaccurate band edge and band gap positions~\cite{Bang2013}. To ensure consistency with the defect-formation-energy formalism, we therefore determine the HSE band edges and band gap from the ionization potential and electron affinity of the pristine supercell (Fig.~\ref{fig:sizeeffectgap})~\cite{Bang2013}. Figure~\ref{fig:sizeeffectgap} compares the band edges obtained from two different methods as a function of supercell size for both the HSE and PBE+U functionals. The 256- and 864-atom supercells correspond to $4 \times 4 \times 4$ and $6\times6\times6$ multiples of the primitive cubic cell, respectively, whereas the 512-atom supercell has a rhombohedral shape. In method~A, the band-edges are readily obtained as the CBM and VBM eigenvalues of the pristine supercell. In method~B, the CBM and VBM are extracted from total energy differences between the neutral pristine supercell and supercells with one extra or one fewer electron, respectively. The two methods yield slightly different CBM and VBM positions. The difference between the energy states for CBM and VBM using methods A and B tends to zero as the value of the electronic smearing width approaches zero. In this work, we have used a Gaussian smearing width of 0.05~eV.

\subsection{Defects simulation\label{Sec:DefectDFT}}

Intrinsic point defects in Cu$_3$N were simulated using a $4 \times 4 \times 4$ supercell with a linear size $L = 4a$, containing 256 atoms in the pristine case. The Brillouin zone of the supercell was sampled at a single $\Gamma$-point, which avoids band filling errors and improves computational efficiency by enabling the use of real-valued linear-algebra routines. For $N\times N\times N$ supercells with even $N$, the R and M points of the primitive Brillouin zone fold onto the supercell $\Gamma$-point, so that the VBM and CBM are correctly captured in a single $\Gamma$-point calculation. This folding does not occur for odd $N$, as shown below, and single-$\Gamma$ calculations with odd-multiple supercells therefore fail to capture the true band-edge states and shallow defects. For this reason, all defect calculations reported here employ even-multiple supercells to ensure reliable sampling of the band edges. Materials with band extrema at \textbf{k}-points located away from the boundary of the first Brillouin zone, such as silicon, may require a distinct \textbf{k}-point for each supercell.

Formation energies and transition levels were computed by first relaxing the internal atomic coordinates with PBE+U and then performing a static HSE total-energy calculation on the PBE+U optimized structure. Charged defects were modeled by adding or removing electrons from the supercell, together with a compensating uniform background charge. In all cases, the supercell lattice parameter $L$ was kept fixed during the structural optimization to avoid spurious lattice and strain effects. This protocol, which combines PBE+U structural relaxation with HSE total energy calculations, substantially reduces the computational cost while preserving quantitative accuracy and qualitative trends for defect energetics.

Following Ref.~\cite{Freysoldt2014}, the formation energy $H(X^q)$ of a defect $X$ with charge $q$ (in units of the elementary charge $e$~\cite{Millikan1913}) as a function of the Fermi level $E_F$ is given by
\begin{multline}\label{FormEnergy}
    H(X^q) = E[X^q] - E[0] + \sum_j n_j \mu_j \\
    + q\mu_e + \Delta E_{\mathrm{PA}} + \Delta E,
\end{multline}
where $E[X^q]$ and $E[0]$ are the total energies of the defect supercell in charge state $q$ and the pristine supercell, respectively. The integers $n_j$ and the chemical potentials $\mu_j$ describe the exchange of atomic species with external reservoirs, with $n_j<0$ ($n_j>0$) for atoms added to (removed from) the supercell. The defect charge $q<0$ ($q>0$) corresponds to electrons added (removed), and $\mu_e$ is the electron chemical potential (Fermi level). It is convenient to express $\mu_e$ as
\[
\mu_e = E_{\mathrm{VBM}} + E_F,
\]
where $E_F$ is measured relative to the VBM of the pristine structure. 

In plane-wave codes such as VASP, single-particle eigenvalues are referenced to the cell-averaged Hartree potential, which includes an arbitrary constant arising from the local parts of the PAW potentials. The potential-alignment term $\Delta E_{\mathrm{PA}}$ [Eq.~\eqref{eq:PA}] corrects for the shift between the energy references of the defect and pristine supercells that results from adding or removing atoms and electrons, which changes the average Hartree potential. In Eq.~\eqref{eq:PA}, the quantity $q$ is interpreted as the number of electrons removed (consistent with Eq.~\eqref{FormEnergy}), while $V$ and $V_0$ denote the corresponding single-particle potential energies. In this convention, $\Delta V = V - V_0$ represents the shift of the potential-energy reference for single-particle levels and must be subtracted from the single-particle energies (e.g., in band diagrams) of the defect supercells. Finally, $\Delta E$ collects all remaining finite-size effects beyond potential alignment, which are discussed below.

To analyze the size dependence of the finite-size correction, we first compute the formation-energy contribution
\begin{equation}
\label{FormEnergycore}
    X(N,q) = \left(E[X^q] - E[0] + \Delta E_{\mathrm{PA}}\right)_{N\times N\times N}
\end{equation}
using $N\times N\times N$ supercells ($N = 4, 6, 8$) with a linear size $L = Na$. These size-scaling calculations are carried out at the PBE+U level, which allows us to treat larger supercells than are feasible with the HSE functional. Note that the VBM also shows a weak dependence on supercell size when only the $\Gamma$-point is used; this effect is treated separately. Defining
\begin{equation}
\label{DeltaH}
 \Delta H(N,q)\equiv X(N,q)-X(4,q),   
\end{equation}
the quantity $\Delta H(N)$ was analyzed as a function of $1/L$ and extrapolated to the $1/L\to0$ limit. The extrapolated value,
\begin{equation}
 \Delta E \equiv \lim_{1/L\to 0}\Delta H(N,q)=H(\infty,q)-X(4,q),   
\end{equation}
was then added to the HSE formation energies obtained for the $4 \times 4 \times 4$ cell, assuming that the residual size error is independent of the functional. This procedure yields formation energies representative of the infinite-size limit while retaining the accuracy of the HSE functional. Our expression for the defect formation energy is
\begin{multline}\label{FormEnergyconverged}
    H(X^q) = \left(E[X^q] - E[\textrm{0}]+ \Delta E_{PA}\right)_{4\times 4\times 4}^{\textrm{HSE}} +  \Delta E+  \\  +\sum_{j}{n_j \mu_j} + q\left(E_{\textrm{VBM}}\right)_{4\times 4\times 4}^{\textrm{HSE}} + q E_F  .
\end{multline}

\begin{figure}[tbh!]
    \centering   \includegraphics[width=0.9\linewidth]{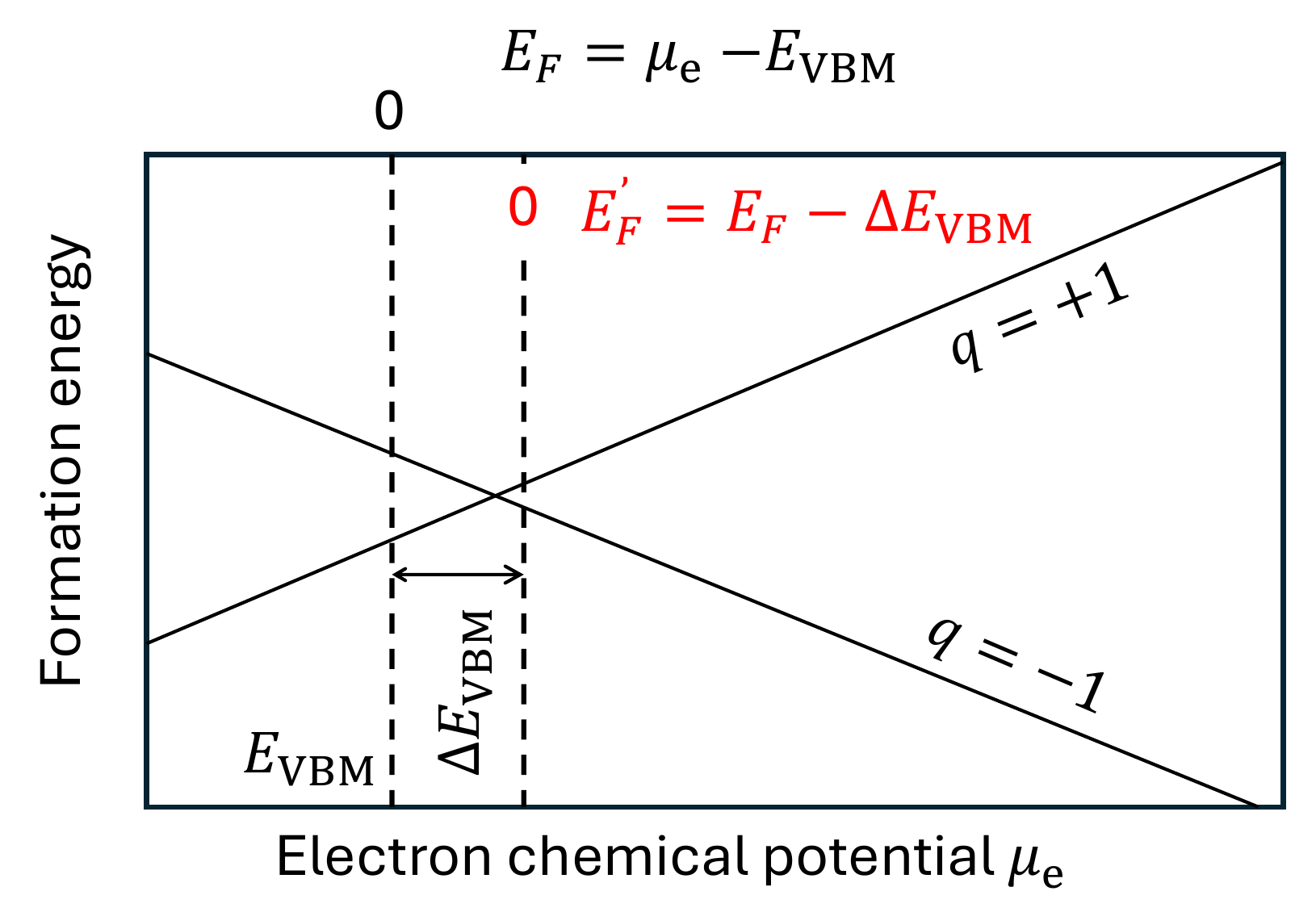}
    \caption{Scheme of formation energy as a function of the electron chemical potential $\mu_e$. The definition of the valence band maximum has an effect on the description of the defect properties. }
    \label{fig:scheme}
\end{figure}

In Eq.  \eqref{FormEnergyconverged}, $\left(E_{\textrm{VBM}}\right)_{4\times 4\times 4}^{\textrm{HSE}}$ denotes the generalized Kohn–Sham eigenvalue of the highest occupied crystal orbital of the pristine supercell. Accordingly, $E_F$ is the Fermi level measured with respect to this VBM energy. Corrections to the VBM can be included by adding a term $q\,\Delta E_{\textrm{VBM}}$ and modifying the Fermi level definition as $E_F^{\prime} = E_F - \Delta E_{\textrm{VBM}}$, as illustrated in Fig.~\ref{fig:scheme}. Generally, the Fermi level takes values between the VBM and the CBM, i.e., within the forbidden band. The correction $\Delta E_{\textrm{VBM}}$ may change the stable charge states. In the schematic example of Fig.~\ref{fig:scheme}, the charge state $q=+1$ is stable near $E_F\simeq$ 0, but after considering the correction of the VBM, the $q=-1$ charge state is the only stable state. Similarly, corrections to the CBM can modify the stable charge states near the CBM. The quantity $\Delta E_{\textrm{VBM}}$ has three contributions: (1) the definition of the VBM as an ionization energy (see Fig.~\ref{fig:sizeeffectgap}), (2) finite-size effects, and (3) quasiparticle corrections.  The associated finite-size corrections are small (below 0.03~eV) and affect both VBM definitions and both functionals similarly. Quasiparticle corrections beyond the HSE level are not used in this work.

\subsection{Thermodynamic constraints}

Having established the electronic reference energies, we now specify the thermodynamic constraints on the atomic reservoirs. In thermodynamic equilibrium, the chemical potentials of the constituent elements and compounds must satisfy
\begin{equation}
\label{ec:Thermo1}
3\mu_{\mathrm{Cu}}+\mu_{\mathrm{N}}=\mu(\mathrm{Cu}_3\mathrm{N}).
\end{equation}
It is convenient to express the elemental chemical potentials relative to their standard elemental references as
\begin{equation}
\label{ec:Thermo2}
\mu_{A}=\mu_{A}^{0}+\Delta\mu_{A},
\end{equation}
where $\mu_{A}^{0}$ is the chemical potential of the pure element in its stable phase and $\Delta\mu_{A}$ measures the deviation imposed by the growth environment (with the constraint $\Delta\mu_{A}\le 0$). The chemical potential of the compound is given by its total energy per formula unit,
\begin{equation}
\label{ec:Thermo3}
\mu(\mathrm{Cu}_3\mathrm{N}) = 3\mu_{\mathrm{Cu}}^{0}+\mu_{\mathrm{N}}^{0}+\Delta H(\mathrm{Cu}_3\mathrm{N}),
\end{equation}
where $\Delta H(\mathrm{Cu}_3\mathrm{N})$ is the heat of formation. Combining Eqs. 16--18 yields the linear constraint
\begin{equation}
\label{ec:Thermo4}
 3\,\Delta\mu_{\mathrm{Cu}}+\Delta\mu_{\mathrm{N}}=\Delta H(\mathrm{Cu}_3\mathrm{N}),   
\end{equation}
which defines the allowed $(\Delta\mu_{\mathrm{Cu}},\Delta\mu_{\mathrm{N}})$ range used in the defect formation-energy calculations.

The total energies of Cu$_3$N and the competing phases (Cu, N) were computed at the HSE level to define consistent elemental chemical potentials ($\mu_{\Cu}^{0}=-3.632$~eV, $\mu_{\N}^{0}=-5.450$~eV) and ensure phase stability ($\Delta H(\Cu_3\N)=-4.110$ eV). Table~\ref{chem-pot} summarizes the relative elemental chemical potentials for Cu and N at the Cu-rich and N-rich limits. Cu$_3$N is metastable against decomposition into bulk copper and molecular nitrogen. In previous studies of Cu$_3$N defects~\cite{Yee2018, Madhu2026,Fioretti2016}, the formation energies in N-rich condition have been computed by setting the chemical potentials at conventional, rather arbitrary, values, appealing to the concept of activated nitrogen. These chemical potentials~\cite{Yee2018, Madhu2026} are shown in Table~\ref{chem-pot} as the $\N^{\ast}$-rich condition. 
Fioretti et al.~\cite{Fioretti2016} used similar values. In contrast, we have adopted the convention of Ref.~\cite{Caskey2014}, considering $\Delta H ($Cu$_3$N$)$ from bulk copper and atomic (dissociated) nitrogen. The relation between our N-rich and their $\N^{\ast}$-rich condition is given in Table~\ref{chem-pot}. Note that temperature and pressure at the N-rich condition in gas phase are accounted by the function $\Delta\mu_{\N}(T,P)$, which satisfies $\Delta\mu_{\N}=0$ in the limit when $T\rightarrow{}0$~K (see SM~\cite{SI_Cu3N}).

\begin{table}
\centering
\caption{Chemical potentials (eV) of copper and nitrogen relative to the reference states used in the literature and in our work. The N-rich condition corresponds to equilibrium with atomic nitrogen gas at zero temperature. The N$^*$-rich condition has been used in Refs.~\cite{Yee2018, Madhu2026}.}
\begin{tabular}{l c c c c c}
\hline \hline
Chem. pot. & \multicolumn{2}{c}{Refs.~\cite{Yee2018, Madhu2026}} &
\multicolumn{3}{c}{{Our work}} \\
\cmidrule(lr){2-3}
\cmidrule(lr){4-6}
& Cu-rich & N$^*$-rich & Cu-rich & N$^*$-rich & N-rich \\
\hline
$\Delta\mu_{\mathrm{Cu}}$
& +0.00 & $-$0.64 & +0.00 & $-$0.66 & $-$1.37 \\
$^{a}\Delta\mu_{\mathrm{N}}$
& $-$4.05 & $-$2.13 & $-$4.11 & $-$2.13 & +0.00 \\
$^{b}\Delta\mu_{\mathrm{N}^*}$
& +0.72 & +2.64 & +0.66 & +2.64 & +4.77 \\
\hline \hline
\end{tabular}

\vspace{0.5em}
\begin{flushleft}
\footnotesize
$^{a}\Delta\mu_{\mathrm{N}} = \mu_{\mathrm{N}} - E_{\mathrm{N}}$; $^{b}\Delta\mu_{\mathrm{N}^*} = \mu_{\mathrm{N}} - \frac{1}{2}E_{\mathrm{N}_2}$; 
$\Delta\mu_{\mathrm{N}^*} = \Delta\mu_{\mathrm{N}} + E_{\mathrm{N}} - \frac{1}{2}E_{\mathrm{N}_2}
= \Delta\mu_{\mathrm{N}} + 4.77~\mathrm{eV}$
\end{flushleft}
\label{chem-pot}
\end{table}

\section{Results}

\subsection{Analysis of size corrections fitting models}

We discuss size corrections for low-energy intrinsic defects: (i) Cu interstitial (Cu$_{\textrm{i}}$), (ii) N interstitial (N$_{\textrm{i}}$), (iii) Cu vacancy (V$_{\textrm{Cu}}$), and N vacancy (\defe{V}{N}). Figure~\ref{structures} shows the lowest energy structure for the neutral case Cu$_{\textrm{i}}^0$ (b), N$_{\textrm{i}}^0$ (c), V$_{\textrm{Cu}}^0$ (d), and V$_{\textrm{N}}^0$ (e) after performing structural relaxations. The antisite defects Cu$_\mathrm{N}$ and N$_\mathrm{Cu}$ have large formation energies and are therefore not considered in this work~\cite{Yee2018}. In principle, interstitial defects can occupy multiple sites within the Cu$_3$N cavity. However, previous studies have shown that Cu$_{\textrm{i}}$ stabilizes at the center of the cubic $Pm\bar{3}m$ structure, and we therefore initialized the structure in this configuration~\cite{Yee2018,Ji2010,Lu2013}. In contrast, for N$_{\textrm{i}}$, due to the small ion size of N, we consider multiple initial positions, including a tetragonal configuration of N bonded to three Cu ions and N in the corner; a diagonal N$_{\textrm{i}}$ forming a plane with Cu atoms; and N$_{\textrm{i}}$ near the center of the face of the unit cell, exhibiting corner placement towards the center of the cubic cavity. For V$_{\textrm{Cu}}$ and V$_{\textrm{N}}$, we find that removing the Cu and  N ions, respectively, induces negligible perturbation of the pristine structure. 

\begin{figure}[hb!]
\includegraphics[width=0.90\columnwidth]{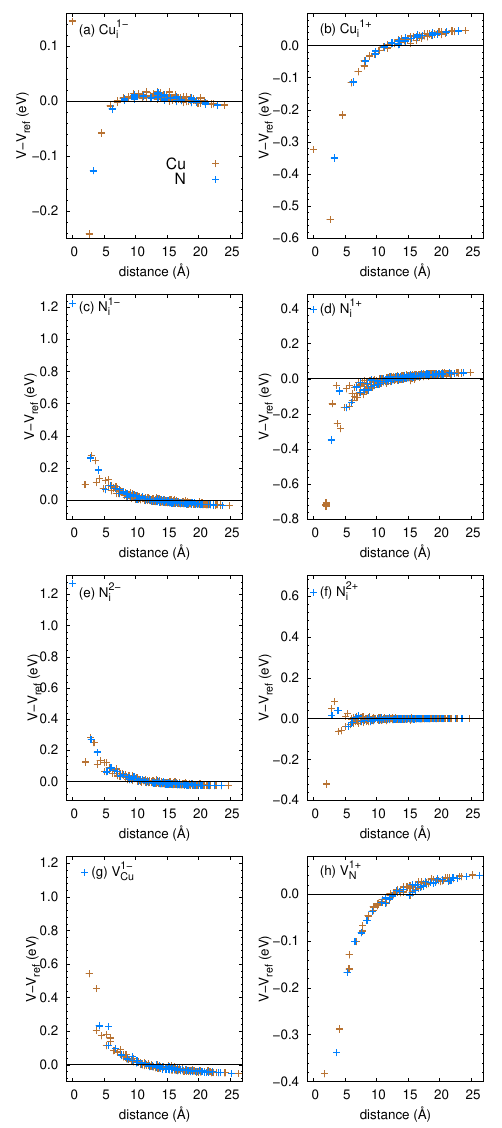}
\caption{The core-level shifts are shown as a function of distance from the defect for: (a) Cu$_{\textrm{i}}^{1-}$, (b) Cu$_{\textrm{i}}^{1+}$, (c) N$_{\textrm{i}}^{1-}$, (d) N$_{\textrm{i}}^{1+}$, (e) N$_{\textrm{i}}^{2-}$, (f) N$_{\textrm{i}}^{2+}$, (g) V$_{\textrm{Cu}}^{1-}$ and (h) \defeq{V}{N}{1+}. All calculations were performed using an $8 \times 8 \times 8$ supercell and the PBE+U functional.}
\label{potentials}
\end{figure}

To assess the degree of structural relaxation around each of the neutral defects with supercell size, we compare the mean atomic displacement around each defect as a function of neighboring atomic shells. Figures S2 and S3 in the SM report, respectively, displacement magnitude versus distance to defect, and mean displacement as a function of neighbor shell, at the PBE+U level, for different supercell sizes. The data are summarized in Table S1 in the SM. We note that for all cases, structural distortions are noticeable beyond the first coordination shell, and the distortion size depends on the defect type and does not follow a single universal trend.

For each charge defect, we computed the potential alignment correction $\Delta E_{PA}$ (see Eq.~\eqref{eq:PA}) for the charge defect energies. Figure~\ref{potentials} shows the potential average values in the $8\times 8 \times8$ defective supercell with respect to the reference values -- those of the pristine supercell -- as a function of the distance to the impurity or the vacancy center. Figures~S4--S7 show the results for $4 \times 4 \times 4$ and $6 \times 6 \times 6$ supercells. The potential alignment was obtained as average values of the electrostatic potential in atomic spheres defined by the DFT code. These averages have a single value for each species, i.e., Cu or N, in the pristine supercell, but they vary across the defective supercell. The potential values are multiplied by $-e$~\cite{Millikan1913}; hence, these values are potential energies in eV units. 

We find that the potential values are well converged to zero at large distance values, suggesting that the $8 \times 8 \times 8$ supercell is large enough to simulate an isolated impurity. As an estimation of the potential alignment correction, we used the value corresponding to the largest distance in the plot, multiplied by the defect charge $q$. Most of the plots in Fig.~\ref{potentials} have the appearance of Coulomb pair potential energies, although they rather represent the interaction with a periodic array of defects imposed by the periodic boundary conditions. The cases of \defeq{Cu}{i}{1-} and \defeq{N}{i}{2+} are anomalous and will be discussed below.

\begin{figure*}[ht]
\includegraphics[width=\textwidth]{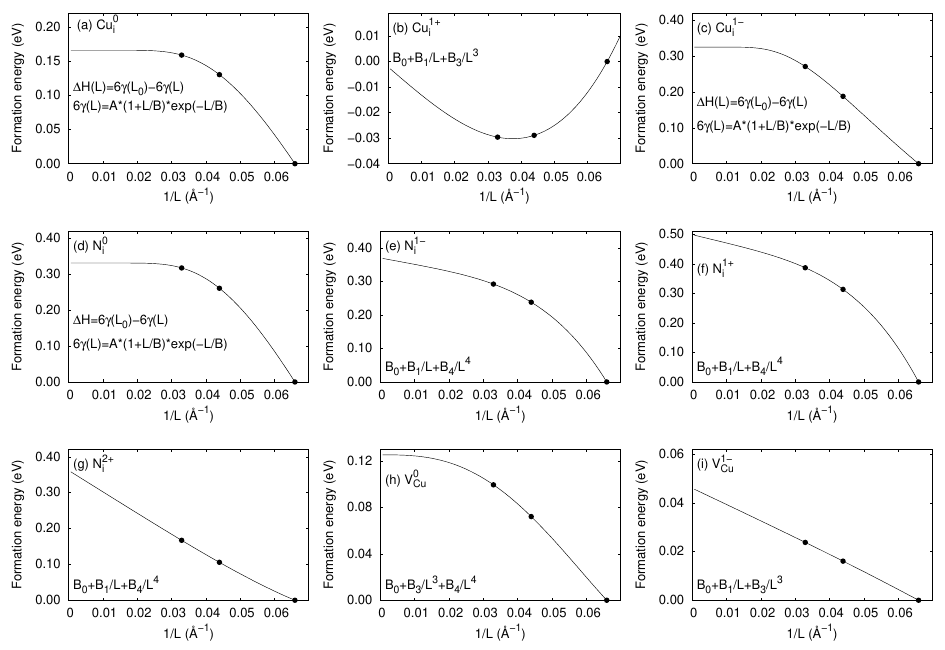}
\caption{Finite–size scaling for defect-formation energies in Cu$_3$N. (a–c) Cu interstitials $\mathrm{Cu_i}^{0,1+,1-}$; (d–g) N interstitials $\mathrm{N_i}^{0,1-,1+,2+}$; (h,i) Cu vacancies $\mathrm{V_{Cu}}^{\,0,1-}$. Symbols are total energies obtained with the PBE+$U$ functional for $N\times N\times N$ supercells with $N=4,6,8$ (256, 864, and 2048 atoms, respectively). }
\label{fitting}
\end{figure*}

\begin{figure}[ht!]
\includegraphics[width=\columnwidth]{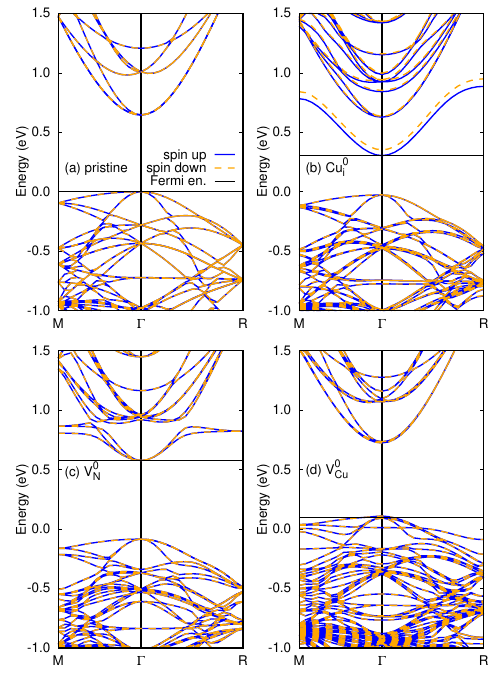}
\caption{Band structures for (a) pristine Cu$_3$N, (b) Cu$_{\textrm{i}}^0$, (c) \defeq{V}{N}{0}, and (d) V$_{\textrm{Cu}}^0$ defects computed in a $4 \times 4 \times 4$ supercell using the PBE+U functional. Energies are with relative to the VBM, and the alignment correction is applied in (b,c,d). As all the calculations were performed using $\Gamma$-point it is possible to see the in gap level introduced in Cu$_{\textrm{i}}^0$ defect. The horizontal black lines represent the Fermi energy for $\Gamma-$point calculation in each case.}
\label{bands}
\end{figure}

Figure~\ref{fitting} presents finite size corrections to defect formation energies, obtained by extrapolating results from $N\times N \times N$ supercells ($N=4, 6, 8$) at the PBE+U level. The best fitting model is shown for each defect and charge state, and the discussion that follows provides support for each model selection. Other fitting models are shown in the SM. Table S2 in the SM shows the data used to obtain the size corrections by means of different fitting models. These data are the formation energy differences (Eq.~\eqref{DeltaH}). Table~\ref{corrections} summarizes the size correction results and highlights the best value used to predict formation energies and defect levels in the last section.

\subsubsection{Interstitial copper defect Cu$_{\textrm{i}}$}

Figure~\ref{fitting} reports finite size corrections to the formation energies of \defe{Cu}{i} for charge states $q=0,\pm 1$. Figure~S8 in the SM includes the \defeq{Cu}{i}{0} formation energy and compares two polynomial forms with the hydrogen-like model introduced earlier, which is the model shown in Fig.~\ref{fitting}(a). Figure~S8 shows an outlier at $1/L = 0.09\,\text{\AA}^{-1}$, corresponding to the formation energy computed with a $3 \times 3 \times 3$ supercell. This outlier demonstrates that odd $N$ supercells can yield anomalous values for the formation energy because they do not include the true VBM and CBM when only the $\Gamma$-point is used. This effect has been observed in the finite-size analysis of charged defects~\cite{Castleton2009}. 

On the other hand, as the net charge is zero, the Makov-Payne monopole correction and the potential alignment do not explain the deviation trend with supercell size. Dipole or quadrupole corrections are also not expected because the impurity atoms are located in a centrosymmetric position. Therefore, we search for alternative fitting functions. 

Figure~\ref{bands} shows the band structure of pristine Cu$_3$N and the neutral defect Cu$_{\textrm{i}}^0$, shown in panels (a) and (b), respectively, using the 4 $\times$ 4 $\times$ 4 supercell and the PBE+U functional in both cases. Since the formation energies are computed using single $\Gamma$-point calculations, only the band energies at $\Gamma$ contribute to the calculated formation energy. Cu$_{\textrm{i}}^0$ exhibits two in-gap bands with opposite spin and similar dispersion. Only one of the two levels at the $\Gamma$-point is occupied, which explains the formation of the $1+$ and $1-$ charge states, as well as the values of the $\epsilon(1+/0)$ and $\epsilon(0/1-)$ transition levels.

Because the formation energies are computed using single $\Gamma$-point calculations, the behavior of the bands at $\Gamma$ determines the accuracy of the fitting model. In the case of Cu$_{\textrm{i}}$, the shift of the $\Gamma$-point energy from the band center is the primary cause of the size effect. The pronounced dispersion indicates interaction between periodic images of the defect. Therefore, using the hydrogen-like fitting model (see Eq.~\eqref{eq:expo}), we obtain a formation energy correction of 0.17~eV. Figure~S8 also shows that fitting the formation energies using polynomial functions, excluding the linear term, yields an extrapolated correction energy that is relatively close to that obtained with the hydrogen-like model.

\begin{figure}[tbh!]
\includegraphics[width=\columnwidth]{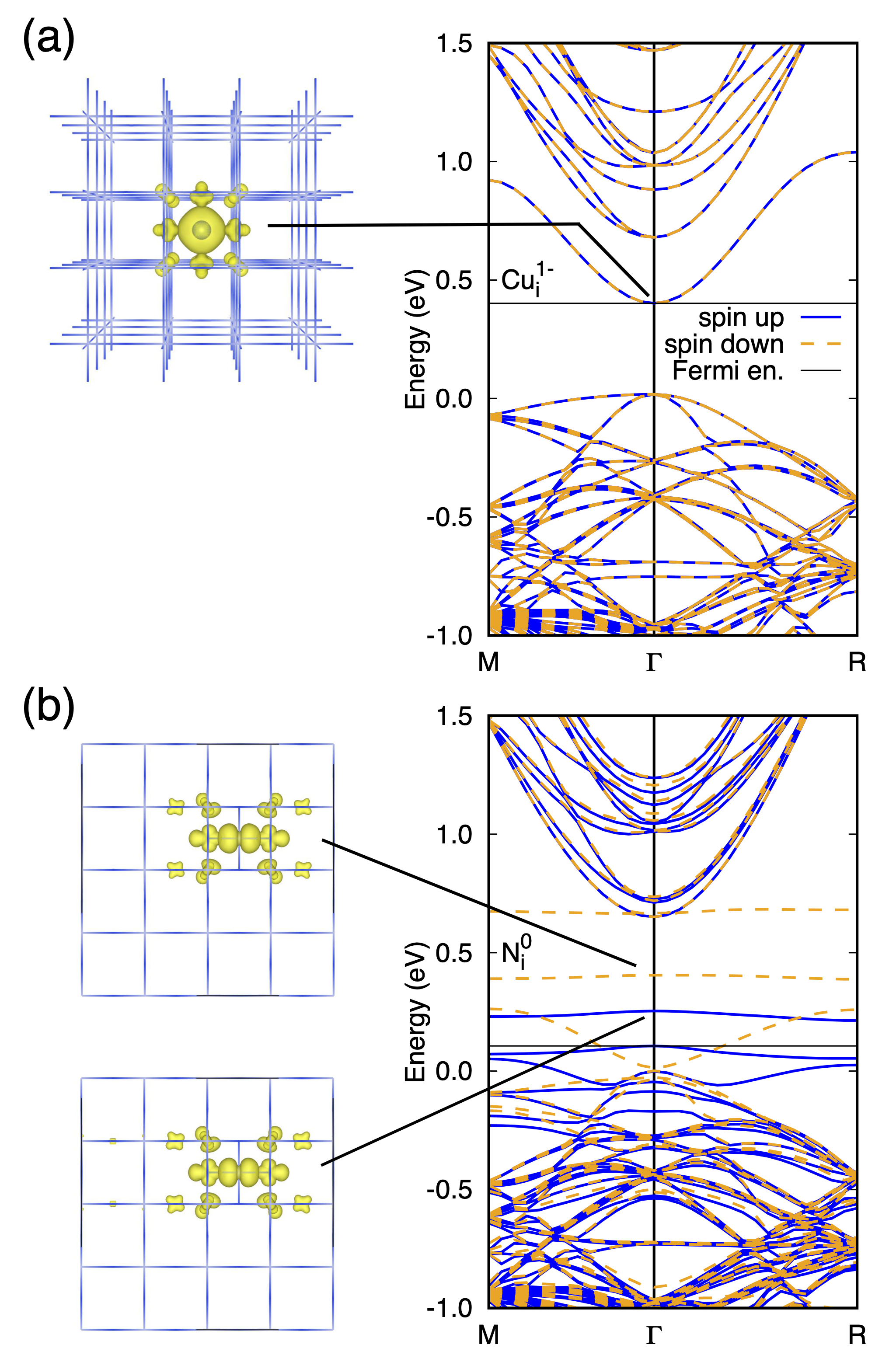}
\caption{(a) Isosurface of the partial charge density of the doubly occupied in-gap state of Cu$_{\textrm{i}}^{1-}$. (b) Partial charge densities of N$_{\textrm{i}}^0$ the impurity states and band edge states. The isosurfaces represented are 0.00112~\AA$^{-3}$. The band energies are relative to the VBM of the pristine material, and include potential alignment corrections.}
\label{charge}
\end{figure}

Additional insight can be obtained by considering the band diagram in the framework of the $6\times 6\times 6$ supercell. In order to reduce the computational cost, the bands were computed for the charge state $q=1-$, which allows a non-spin-polarized calculation. Figure~\ref{charge}(a) shows the bands for the Cu$_{\textrm{i}}^{1-}$ defect, which are nearly identical to the bands of the neutral state but are spin-degenerate. The latter is compared to the band diagram for the $6 \times 6 \times 6$ supercell shown in Fig.~S9 (left panel), computed with only a few \textbf{k}-points to reduce the high computational cost. It can be noted that the in-gap band is narrower in the $6 \times 6 \times 6$ supercell than in the $4\times 4 \times 4$ supercell, but the average band energy is practically the same, and it is very close to the CBM. Figure~\ref{charge}(a) also shows the interstitial atom at the supercell center, with an isosurface of the partial charge density (the square of the wavefunction) of the doubly occupied in-gap state of Cu$_{\textrm{i}}^{1-}$. The decrease in bandwidth upon increasing the supercell size supports the idea that the bandwidth and the deep character of the in-gap level are size effects resulting from the interaction between the interstitial and its periodic replicas, and that Cu$_{\textrm{i}}$ is a donor with shallow-like electronic character near the CBM. As will be shown below, the apparent deep character of the $(0/1-)$ transition level is a size effect.

Figure~S11 shows the formation energies and the fitted curves with the extrapolated values for Cu$_{\textrm{i}}^{1-}$. According to the previous discussion, the impurity model was chosen as the best fitting model, providing a size correction of 0.33~eV. The enlarged value compared to the case of \defeq{Cu}{i}{0} is the result of the double occupation of the in-gap level. In principle, the Makov-Payne model could be applied to \defeq{Cu}{i}{1-}, and this fit is also shown in Fig.~S11, providing a size correction of 0.50~eV. However, the behavior  of the core-averaged potential shown in Fig.~\ref{potentials}(a) suggests the absence of a negatively point-like charge associated with the defect, which also agrees with the dispersive character of the in-gap band. Hence, we state that the Makov-Payne model is not appropriate for \defeq{Cu}{i}{1-}.

\begin{table}[tbh!]
\caption{Summary of finite-size corrections obtained by extrapolating PBE+U formation energies of $N\times N\times N$ supercells with $N = 4, 6$ and $8$. Formation energies for charged states include potential alignment corrections. We consider the following correction methods for $\Delta E$: rigid shift by Eq.~\eqref{DeltaH} with $N=8$ , polynomial scaling with and without $1/L$ term for charged and neutral defects, respectively (see Eq.~\eqref{models}) and hydrogen-like impurity models (Eq.~\eqref{eq:expo}). AN denote anomalous cases discussed in the text. NA refer to cases where the model is not applicable.}
\begin{tabular}{llrrr}
\hline \hline 
Defect            & q & Rigid shift & Polynomial scaling & Impurity model  \\
\hline
Cu$_{\textrm{i}}$  &$0$  & 0.16  & 0.18   & \textbf{0.17}  \\
                   &$1-$ & 0.27  & 0.50   & \textbf{0.33} \\
                   &$1+$ & -0.03 & \textbf{0.00}  & -0.05 \\
N$_{\textrm{i}}$   &$0$  & 0.32  & 0.36  & \textbf{0.33} \\
                   &$1-$ & 0.29  & \textbf{0.37}     & 0.31 \\
                   &$1+$ & 0.39  & \textbf{0.50}     & 0.41\\
                   &$2-$ & 0.25  &            AN     & AN \\
                   &$2+$ & 0.17  & \textbf{0.36}     & 0.23 \\
                   &$3-$ & 0.19  & \textbf{0.25}     & 0.20 \\
V$_{\textrm{Cu}}$  &$0$  & 0.10  & \textbf{0.13}     & 0.11 \\
                   &$1-$ & 0.02  & \textbf{0.05}     & 0.03 \\
V$_{\textrm{N}}$   &$0$  & 0.13  & \textbf{0.16}     & NA \\
                   &$1+$ & 0.09  & \textbf{0.14}     & NA \\
                   &$2+$ & \textbf{0.07}    & AN     & NA \\
                   &$3+$ & \textbf{0.14}    & {0.14} & NA \\
\hline \hline
\end{tabular}
\label{corrections}
\end{table}

\begin{figure*}[bt!]
\includegraphics[width=\linewidth]{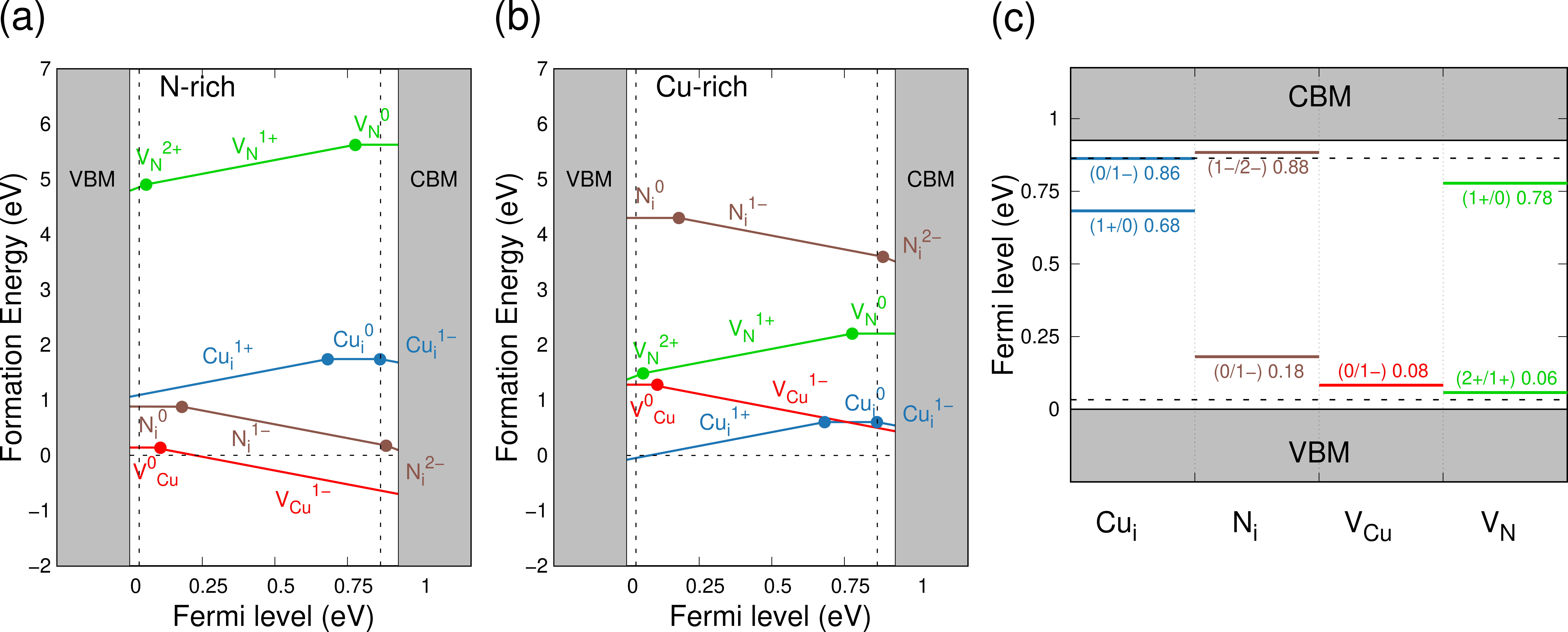}
\caption{
Formation energies of Cu$_\mathrm{i}$, N$_\mathrm{i}$, V$_\mathrm{Cu}$, and V$_\mathrm{N}$ point defects in Cu$_3$N as a function of the Fermi level. For (a) N-rich and (b) Cu-rich conditions we assume quasi-equilibrium with Cu$_3$N, Cu-metal, and atomic N. Panel (c) shows the thermodynamic charge transition levels computed including the finite-size corrections obtained from the fitting models introduced in this work; the formation and transition-level energies are labeled in eV. The valence-band maximum (VBM) and conduction-band minimum (CBM), shown as gray blocks, correspond to the HSE generalized Kohn-Sham eigenvalues. The vertical dashed lines mark the band edges obtained from the electron ionization and affinity energies (method B in Fig.~\protect\ref{fig:sizeeffectgap}), including the size-effect correction, at 0.03 eV for the VBM and 0.86 eV for the CBM.}
\label{defects}
\end{figure*}

For the positive charge state, \defeq{Cu}{i}{1+}, the monopole Makov-Payne correction may be appropriate, as the interstitial atom has lost its outer 4$s^1$ electron and behaves as a positively charged point defect. The trend observed in the core-averaged electrostatic potential shown in Fig.~\ref{potentials}(b) supports this interpretation. Figure~S10 presents several fits using functions that include a monopole-type linear term along with a higher-order $1/L$ term.  The choice for the best model is uncertain due to the fact that the data fall outside the linear convergence regime. However, in all cases, the   extrapolated corrections are very small  (0.02, 0.0, and -0.01~eV) and are well below the accuracy of DFT calculations.  We adopt the intermediate value of 0.0~eV, obtained from fitting the function $B_0 + B_1/L + B_3/L^3$, which is consistent with the Makov-Payne formalism. This fitting is shown in Fig.~\ref{fitting}(b). The fact that $B_1<0$ is consistent with the Makov-Payne model, as discussed in Sec.~\ref{sec:intro}.

\subsubsection{Interstitial nitrogen defect N$_{\textrm{i}}$}

The most stable position of interstitial nitrogen is located near the center of the unit cell face, coordinated with four Cu atoms, as shown in Fig.~\ref{structures}(c). The vertical displacement of the defect from the cell face varies with the defect charge. The band structure of N$_\textrm{i}^0$ exhibits several in-gap bands, the character of which is clarified by the partial charge density at the $\Gamma$-point shown in Fig.~\ref{charge}(b) for the $4 \times 4 \times 4$ supercell. The corresponding squared wave functions are localized around the interstitial site and the neighboring atoms.

For N$_{\textrm{i}}^0$, Fig.~S12 shows the formation energies of the neutral charge state and its fits with two models. First, a model that considers dipole-dipole interactions ($B_3/L^3$) plus residual interactions, and second, the hydrogenic impurity model used for Cu$_{\textrm{i}}^0$. Both models provide nearly the same correction; we chose the impurity model, also in Fig.~\ref{fitting}(d), which has a correction of 0.33~eV. This rather large size correction is surprising. For \defeq{N}{i}{0}, the calculated electronic structure shows a clear size dependence. In the 4×4×4 supercell, the band diagram of Fig.~\ref{charge}(b) shows a spin down dispersive band crossing the Fermi energy and several non-dispersive bands near the Fermi energy, whereas in larger supercells (see Fig.~S9) the bandwidth is reduced and the relative band ordering changes. This indicates that the residual size effect arises from defect-band dispersion, wavefunction overlap, and possibly dipolar interactions. Together with the \defeq{Cu}{i}{0}  defect discussed earlier, this underlines the need to apply size corrections to the neutral defect formation energy.

For N$_{\textrm{i}}^{1-}$, core average potentials and band diagrams suggest that the monopole correction must be present in the fitting model. However, the interaction is not isotropic, as seen in Fig.~\ref{potentials}(d), where the core potential energies depend not only on the distance to the impurity. Also, the orbitals have different symmetries. Among the $B_0 + B_1 / L + B_n / L^n$ models, the case with $n = 4$ is the only one that exhibits a dominant monopole character at large supercell sizes (see Fig.~S13). We therefore select the $B_0 + B_1/L + B_{4}/ L^4$ model, also shown in Fig.~\ref{fitting}(e), which yields a correction of 0.37~eV. Most of the alternative models produce corrections close to 0.35~eV, with the exception of the $B_0 + B_1 / L + B_2 / L^2$ model, which deviates significantly.

Figure~S14 shows the formation energies and the fitted curves with the extrapolated values for N$_{\textrm{i}}^{1+}$. Similar to the previous case N$_{\textrm{i}}^{1-}$, we choose the $B_0+B_1/L+B_4/L^4$ model (also in Fig.~\ref{fitting}(f)) that gives a correction of 0.50~eV. For N$_{\textrm{i}}^{2+}$, the extrapolation of total energy corrections is given in Fig.~S15. Again, we choose the $B_0+B_1/L+B_4/L^4$ model (also in Fig.~\ref{fitting}(g)) that gives a correction of 0.36~eV. For N$_{\textrm{i}}^{2-}$, Fig.~S16 shows the least squares fits for the models considered. In this case, the size dependence is anomalous, and the extrapolated energies are unreliable. Note that the dominant monopole correction should be given by $B_1<0$, but all models give a correction with the opposite sign. Hence, we propose the correction as the difference between the $8 \times 8 \times 8$ and the $4 \times 4 \times 4$ calculations, i.e., 0.25~eV. The defect \defeq{N}{i}{3-} has two electrons at the CBM, causing it to be unstable for all values of the Fermi level. Figure~S17 shows the size dependence of the formation energy. The $B_0+B_1/L+B_4/L^4$ model is the best behaved, giving a size correction of 0.25~eV.

\subsubsection{Vacancy of copper V$_{\textrm{Cu}}$}

The neutral V$_\textrm{Cu}^0$ defect exhibits negligible lattice distortion, with Cu–N bond lengths of 1.90~{\AA} varying by no more than 0.03~{\AA}. Its band structure does not show any in-gap states (see Fig.~\ref{bands}(d)). The neutral configuration is non-magnetic within the PBE+U formalism. A spin-polarized calculation with a total spin of 1 yields an energy that is 0.01~eV higher than that of the spin-unpolarized case. There is a delocalized hole state at the VBM, which allows the $1-$ charge state to adopt a closed-shell configuration and lowers its formation energy.

Figure~S18 shows several fitting models applied to V$_\textrm{Cu}^0$, excluding monopole corrections. We select the $B_0 + B_3 / L^3 + B_4 / L^4$ model, which yields a size correction of 0.13~eV. For the charged state V$_\textrm{Cu}^{1-}$, the least-squares fits are presented in Fig.~S19. The Makov-Payne function model $B_0+B_1/L+B_3/L^3$ fits the size dependence, providing a size correction 0.05~eV. Other linear models, i.e., $B_0+B_1/L+B_3/L^2$ and $B_0+B_1/L+B_3/L^4$, give the same correction, showing that the linear term is dominant.

\subsubsection{Vacancy of nitrogen V$_\textrm{N}$}

The nitrogen vacancy is shown in Fig.~\ref{structures}(e). In the neutral state, the Cu--Cu distance changes from 2.683~\AA{} far from the vacancy to 2.489~\AA{} around the vacancy. The neighboring Cu atoms move 0.137~\AA{} towards the vacancy center, and their bond length with the next N atoms increases to 1.953~\AA{}, a variation of 0.056~\AA{} from 1.897~\AA{} far from the vacancy. The neutral charge state \defeq{V}{N}{0} has zero spin and does not create in-gap levels (see Fig.~\ref{bands}(c)), despite the fact that the highest occupied level (with sixfold degeneracy) is occupied by a single  electron. The state \defeq{V}{N}{1+} has a closed shell configuration and a band diagram almost identical to the band diagram of \defeq{V}{N}{0} (see Fig.~S20). In both cases, the band diagrams exhibit band-like dispersion at the CBM and non-dispersive bands above the CBM. The projected density of states (PDOS) of \defeq{V}{N}{0} (Fig.~S21(a)) shows that the threefold-degenerate level at 0.95~eV projects strongly onto the nearest-neighbor Cu atoms, with a smaller but appreciable contribution from the second-neighbor N atoms, and decays rapidly for farther shells. In contrast, the PDOS of the states at 0.58~eV on the nearest-neighbor Cu atoms is much smaller than at 0.95~eV and only moderately larger than on farther Cu shells, with negligible contribution from N atoms. Consistently, the partial charge densities (Figs.~S21(b) and S21(c)) show that the 0.95~eV states are localized around the vacancy, whereas the 0.58~eV states extend over the whole supercell. Therefore, the feature around 0.95 eV is associated with unoccupied localized vacancy levels, while the dispersive band-like conduction state at 0.58 eV is interpreted as a perturbed host state~\cite{Freysoldt2014,Lany2008EMP}. An HSE band-structure calculation shows the same behavior, as shown in Fig.~S20.

Figures S22--S25 show the fitting curves of the formation energies of \defeq{V}{N}{q} for $q=0,1,2,3$. For the neutral state, we have chosen the model with only cubic and quartic terms, giving a size correction of 0.16~eV. For \defeq{V}{N}{1+}, the linear-cubic model gives a size correction of 0.14~eV. For \defeq{V}{N}{2+,3+}, the linear-cubic model gives $B_1>0$, which contradicts the Makov-Payne model. Therefore, we have chosen the size correction as the difference between the formation energies of $8\times 8\times 8$ and $4\times 4 \times 4$ results. Let us note that the cubic-quartic model gives almost the same corrections.

\subsection{Formation energies and defect level properties of Cu$_3$N}

Having computed the finite-size corrections, we next evaluate the formation energies and defect levels. Figures~\ref{defects}(a) and (b) show the formation energies of Cu$_3$N as a function of the Fermi level for N-rich and Cu-rich conditions, respectively. The results incorporate the finite-size corrections; the adopted values are highlighted in boldface in Table~\ref{corrections}. Under N-rich conditions, the nitrogen vacancy V$_{\textrm{N}}$ also has a relatively high formation energy and is thus unlikely to influence the bulk electrical properties. Its formation energy decreases under Cu-rich conditions, but it is still larger than for \defe{V}{Cu} and \defe{Cu}{i}. Therefore, V$_{\textrm{N}}$ does not act as an effective source of free carriers.

Figure~\ref{defects}(c) summarizes the transition levels. The three low-energy defects (Cu$_\mathrm{i}$, N$_\mathrm{i}$, and V$_\mathrm{Cu}$) exhibit distinct behavior: Cu$_\mathrm{i}$ acts as a donor with a (1+/0) transition level within $\lesssim 0.5$~eV of the CBM; but it is also an acceptor with the $(0/1-)$ transition level resonant with the CBM. V$_\mathrm{Cu}$ behaves as a shallow acceptor with a $(0/1-)$ level near the VBM; whereas N$_\mathrm{i}$ forms deeper and overlapping localized electronic states, in line with previous reports~\cite{Peng2013,Yee2018}. Consistent with the charge-neutrality analysis, Cu$_3$N is $p$-type under N-rich conditions (stabilization of V$_\mathrm{Cu}^{1-}$ and N$_\mathrm{i}^{1-}$) and $n$-type under Cu-rich conditions (stabilization of Cu$_\mathrm{i}^{1+}$).

\begin{figure}
    \centering
    \includegraphics[width=\linewidth]{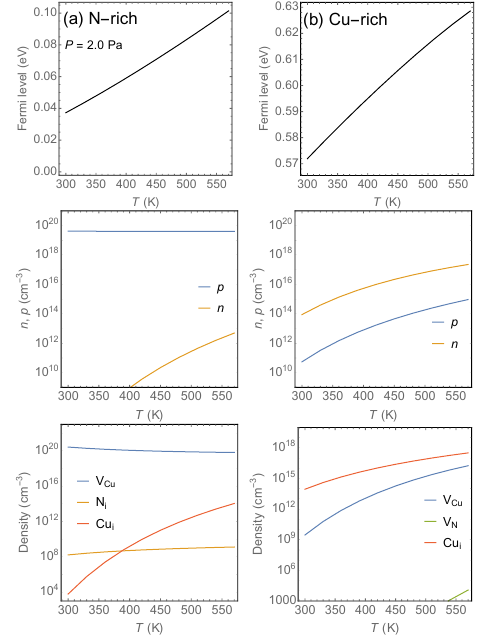}
    \caption{Fermi level, carrier density and defect density as a function of temperature for (a) N-rich and (b) Cu-rich conditions, as determined by solving the charge neutrality equation. The reference for the Fermi level is the same as in Fig.~\protect\ref{defects}. The N-rich condition includes $\Delta \mu_N(T,P)$, with $P=2.0$~Pa.}
    \label{fig:cne}
\end{figure}

To analyze the $n$ or $p$ character of the material, we have determined the Fermi level by solving the charge-neutrality equation (CNE)~\cite{Freysoldt2014} as a function of temperature ($E_F=E_F(T)$), as described in the SM. For N-rich conditions, the low-energy defects are exclusively the acceptor defects, \defe{N}{i} and \defe{V}{Cu}. Therefore, charge neutrality can be achieved between the holes and the negatively charged defects, suggesting a Fermi level between the transition level and the VBM, or even below the VBM. The numerical solution of the CNE, shown in Fig.~\ref{fig:cne}(a), gives values between 0.04 and 0.10~eV for temperatures between 300~K and 570~K, respectively. The defect density, assumed to be in thermal equilibrium, is dominated by \defe{V}{Cu} in a concentration of $10^{20}$ cm$^{-3}$. In N-rich conditions, the hole density achieves a near constant value of $p \sim 4 \times 10^{19}$ cm$^{-3}$. Note that such order of carrier concentration ($1.43 \times 10^{20}$ cm$^{-3}$) has been reported for Cu$_3$N~\cite{Chen2022, Wen2025}. However, those results should be regarded with caution. Our calculations show that for $p = 1.43 \times 10^{20}$ cm$^{-3}$, the Fermi level must be 8 meV below the VBM. This represents an extreme case of heavy self-doping, as well as instability of the material, but it illustrates that strong $p$-character can be achieved under N-rich conditions. Under Cu-rich conditions, \defeq{V}{Cu}{0}, \defeq{V}{Cu}{1-}, and \defeq{Cu}{i}{1+} have the lowest formation energies, which intersect in the range of 0.68-0.76 eV, pushing the Fermi level toward this range. By solving the CNE, we obtain Fermi level values between 0.57 and 0.63~eV, as shown in Fig.~\ref{fig:cne}(b). As expected, the defect densities of  \defe{Cu}{i} and \defe{V}{Cu} prevail, with values as large as $10^{17} \text{ cm}^{-3}$ for \defe{Cu}{i} at high temperatures. The electron density, $n$, takes values between $10^{14}$ and $10^{17}$ cm$^{-3}$, which is significantly smaller than the hole concentration, $p$, under N-rich conditions. This stems from higher defect formation energies, leading to lower defect concentrations, along with the compensating effect of \defe{V}{Cu}.

\section{Conclusions}

In summary, we have performed DFT total energy calculations of formation energies and transition levels for intrinsic point defects in Cu$_3$N.  The PBE+U method was employed to perform structural relaxations, while the HSE method was used to improve the total energies. The formation energies were corrected for potential alignment and size effects. For each defect and each charge state, the size corrections were obtained by extrapolating the defect formation energy  computed with supercells of 256, 864, and 2048 atoms (pristine material). The extrapolation was obtained using a fitting model, which was selected for each defect.

The computed formation energies show that V$_{\mathrm{Cu}}$ acts as an acceptor near the VBM; Cu$_\mathrm{i}$ behaves as donor and resonant acceptor; and N$_\mathrm{i}$ is associated with localized deep electronic states, consistent with prior reports. \defe{V}{N} vacancies are present in negligible concentrations and lack single-electron energy levels within the band gap; therefore, these do not influence optical or electrical properties. Our charge neutrality analysis yields $p$-type behavior under N-rich conditions due to the stabilization of V$_{\mathrm{Cu}}^{1-}$ at $E_F \sim$ 0.08~eV, and $n$-type behavior under Cu-rich conditions due to the stabilization of Cu$_\mathrm{i}^{1+}$ at $E_F \sim$ 0.68~eV.

We analyze explicit finite-size trends across a sequence of large supercells and determine which functional form best describes each defect according to its physical character. This allows us to identify cases in which conventional electrostatic scaling is adequate, as well as cases in which shallow character, defect-band dispersion, wavefunction overlap, or residual short-range interactions require more flexible defect-specific models.

Our work emphasizes that no single finite-size correction method can be applied across all defect types. Different defects and defect states require different size correction methods or extrapolation strategies. For Cu$_3$N, we find that singly charged vacancies are accurately captured by Makov-Payne-type scaling ({$1/L + 1/L^{3}$}), while interstitial defects with shallow or weakly localized electronic character are better described by a hydrogenic impurity model that accounts for defect-band dispersion. Additionally, residual trends for neutral or weakly localized defects are best modeled by higher-order polynomial fits in $1/L^{3}$ and $1/L^{4}$.

Overall, our results support an intrinsic $p$–$n$ doping behavior of Cu$_3$N under realistic chemical potentials and provide a practical, defect-specific protocol for assessing and correcting finite-size effects in semiconductors hosting both shallow and deep defects. Our strategy yields dilute-limit energetics at a practical computational cost and offers practical guidance for selecting correction schemes when defect wavefunctions span multiple length scales.

\begin{acknowledgments}
A.M.R. acknowledges financial support from ANID Fondecyt Postdoctoral grant number 3220460. S.E.R.-L. acknowledges financial support from ANID Fondecyt Regular grant number 1260824. A.M.R. and S.E.R.-L. conducted theoretical work at The Molecular Foundry, which is supported by the Office of Science, Office of Basic Energy Sciences of the U.S. Department of Energy under contract No. DE-AC02-05CH11231. E.M.-P. thanks the computing time provided by the \emph{Servicio de Supercomputación de la Universidad de Granada} (https://supercomputacion.ugr.es). Powered@NLHPC: This research was partially supported by the supercomputing infrastructure of the NLHPC (CCSS210001).

The data that support the findings of this article are openly available at Ref.~\cite{Cu3N_zenodo}.

\end{acknowledgments}

\section*{Author Contributions}

A.M.R.: formal analysis, funding acquisition, investigation, 
visualization, and writing the original draft. S.E.R.-L.: 
conceptualization, funding acquisition, resources, visualization, 
and supervision. E.M.-P.: data curation, formal analysis, 
methodology, resources, and visualization. All authors contributed 
to the investigation and the writing, review, and editing of the final manuscript.

%

\clearpage
\onecolumngrid
\renewcommand\thefigure{S\arabic{figure}}
\renewcommand\thetable{S\arabic{table}}
\renewcommand\theequation{S\arabic{equation}}
\renewcommand\thesection{S\arabic{section}}
\renewcommand\thesubsection{S\arabic{section}.\arabic{subsection}}
\renewcommand\thesubsubsection{S\arabic{section}.\arabic{subsection}.\arabic{subsubsection}}
\renewcommand\thepage{S\arabic{page}}
\setcounter{page}{1}
\setcounter{figure}{0}
\setcounter{table}{0}
\setcounter{equation}{0}
\setcounter{section}{0}

\newgeometry{vmargin={1.5cm,1.5cm}, hmargin={2cm,2cm}}



\begin{center}
\LARGE{\textbf{Supplemental material\\First-principles finite-size correction schemes for  point defects of Cu$_3$N}}
\\
\vspace{0.5cm}
\large{Abdul M. Reyes$^{(1,2)}$,  Sebastian E. Reyes-Lillo$^{(3)}$, and Eduardo Menéndez-Proupin$^{(4)}$ 
\\
\vspace{0.4cm}
\small
(1) Universidad Nacional de Colombia, Sede Tumaco, Tumaco, Nariño 764501, Colombia\\
\small
(2) Departamento de Física, Grupo de Óptica Cuántica y Centro de Excelencia en Nuevos Materiales (CENM),\\
\small
Universidad del Valle, A.A. 25360, Santiago de Cali, Colombia \\
\small
(3) Departamento de F\'isica y Astronom\'ia, Universidad Andres Bello, Santiago 837-0136, Chile  \\
\small
(4) Departamento de Física Aplicada I, Escuela Politécnica Superior, Universidad de Sevilla, Seville E-41011, Spain
}

\end{center}


\newpage

\section{Atomic distances and formation energies}

\begin{table}[H]
\addcontentsline{toc}{table}{Table S1}
\centering
\caption{For different supercell sizes, and for each neutral defect, we summarize the mean distance from the defect to the $n$-neighbor shell, $d$ (\AA{}), and the mean atomic displacement of that shell, $\langle |\Delta \mathbf{r}| \rangle$ (\AA{}), as a function of the shell index $n$. Shells are constructed by sorting all atoms by their distance to the defect and grouping them sequentially: starting from the closest atom not yet placed in a shell, every subsequent atom within 0.15~\AA{} of it is added to the same shell; $d$ is the resulting group-averaged distance.}
\label{tab:SY-888}
\begin{tabular}{c c c c c c}
\hline
Supercell & Shell &
Cu$_i$ $d$  / $\langle|\Delta\mathbf{r}|\rangle$  &
N$_i$ $d$ / $\langle|\Delta\mathbf{r}|\rangle$  &
V$_{\mathrm{Cu}}$ $d$  / $\langle|\Delta\mathbf{r}|\rangle$  &
V$_{\mathrm{N}}$ $d$  / $\langle|\Delta\mathbf{r}|\rangle$  \\
\midrule
$8 \times 8 \times 8$ & 1 & 2.68 / 0.0348 & 2.02 / 0.0414 & 1.90 / 0.0557 & 1.90 / 0.1378 \\
& 2 & 3.29 / 0.0169 & 2.77 / 0.0784 & 2.68 / 0.0456 & 3.79 / 0.0830 \\
& 3 & 4.65 / 0.0080 & 2.94 / 0.0372 & 3.79 / 0.0092 & 4.24 / 0.0614 \\
& 4 & 6.00 / 0.0032 & 3.68 / 0.0217 & 4.24 / 0.0185 & 5.37 / 0.0161 \\
& 5 & 6.29 / 0.0008 & 4.09 / 0.0059 & 4.65 / 0.0076 & 5.69 / 0.0116 \\
& 6 & 7.10 / 0.0012 & 4.30 / 0.0338 & 5.37 / 0.0066 & 6.57 / 0.0008 \\
\hline
$6 \times 6 \times 6$&1 & 2.68 / 0.0311 & 2.02 / 0.0419 & 1.90 / 0.0568 & 1.90 / 0.1370 \\
&2 & 3.29 / 0.0173 & 2.77 / 0.0788 & 2.68 / 0.0531 & 3.79 / 0.0809 \\
&3 & 4.64 / 0.0060 & 2.94 / 0.0359 & 3.79 / 0.0135 & 4.24 / 0.0632 \\
&4 & 6.00 / 0.0026 & 3.68 / 0.0240 & 4.24 / 0.0201 & 5.37 / 0.0149 \\
&5 & 6.29 / 0.0020 & 4.10 / 0.0053 & 4.65 / 0.0142 & 5.69 / 0.0111 \\
&6 & 7.10 / 0.0018 & 4.30 / 0.0338 & 5.37 / 0.0141 & 6.57 / 0.0007 \\
\hline
$4 \times 4 \times 4$ &1 & 2.68 / 0.0301 & 2.01 / 0.0418 & 1.90 / 0.0533 & 1.90 / 0.1320 \\
&2 & 3.29 / 0.0168 & 2.77 / 0.0750 & 2.68 / 0.0521 & 3.79 / 0.0661 \\
&3 & 4.64 / 0.0062 & 2.95 / 0.0433 & 3.79 / 0.0073 & 4.24 / 0.0566 \\
&4 & 6.00 / 0.0046 & 3.68 / 0.0283 & 4.24 / 0.0167 & 5.37 / 0.0124 \\
&5 & 6.29 / 0.0023 & 4.11 / 0.0096 & 4.65 / 0.0143 & 5.69 / 0.0076 \\
&6 & 7.10 / 0.0026 & 4.30 / 0.0289 & 5.37 / 0.0140 & 6.57 / 0.0016 \\
\hline
\end{tabular}
\end{table}

\begin{table}[H]
\addcontentsline{toc}{table}{Table S2}
\centering
\caption{Formation energies for Cu$_i$, N$_i$, V$_\textrm{Cu}$ and V$_\textrm{N}$ in different charge states. Formation energies are computed using PBE+U and $N \times N \times N$ supercells with $N = 4, 6, 8$, relative to the $4 \times 4 \times 4$ supercell. The Fermi level is set at the valence band maximum of the $4 \times 4 \times 4$ supercell. Charged state energies include potential alignment corrections.}
\begin{tabular}{lrcccccc}
\hline
Cu$_i$ &q & 0 & 1+&  $1-$ &&&\\
&N & & & &&\\
\hline
&4 &0.000 &0.0000  &0.0000 &&&\\
&6 &0.130 &-0.0289 &0.1877 &&&\\
&8 &0.159 &-0.0296 &0.2709 &&&\\
\hline
N$_i$ &q & 0 & 1+&  $1-$ & 2+ &  $2-$ & $3-$\\
&N & & & & & &\\
\hline 
&4 &0.0000 &0.0000 &0.0000 &0.0000 &0.0000 & 0.0000\\
&6 &0.2613 &0.3134 &0.2383 &0.1057 &0.4128 & 0.1533\\
&8 &0.3179 &0.3866 &0.2923 &0.1671 &0.2533 & 0.1898 \\
\hline 
V$_\textrm{Cu}$ & q & 0 & $1-$&  1+ & & & \\
&N & & & & & &\\
\hline
&4 &0.0000      &0.0000      & &&& \\
&6 &0.0722 &0.0161 & &&& \\
&8 &0.0996 &0.0238 & &&& \\

\hline 
V$_\textrm{N}$ & q & 0 & 1+&  2+& 3+& \\
&N & & & & & \\
\hline
&4 &0.0000 &0.0000 &0.0000 &0.0000 & \\
&6 &0.0918 &0.0699 &0.0687 &0.1157 & \\
&8 &0.1253 &0.0934 &0.0741 &0.1355 & \\
\hline
\end{tabular}
\end{table}

\newpage

\section{Band structures, atomic distances, potential alignments and finite-size corrections}

\begin{figure}[H]
\addcontentsline{toc}{figure}{Figure S1}
\centering
\includegraphics[width=0.7\linewidth]{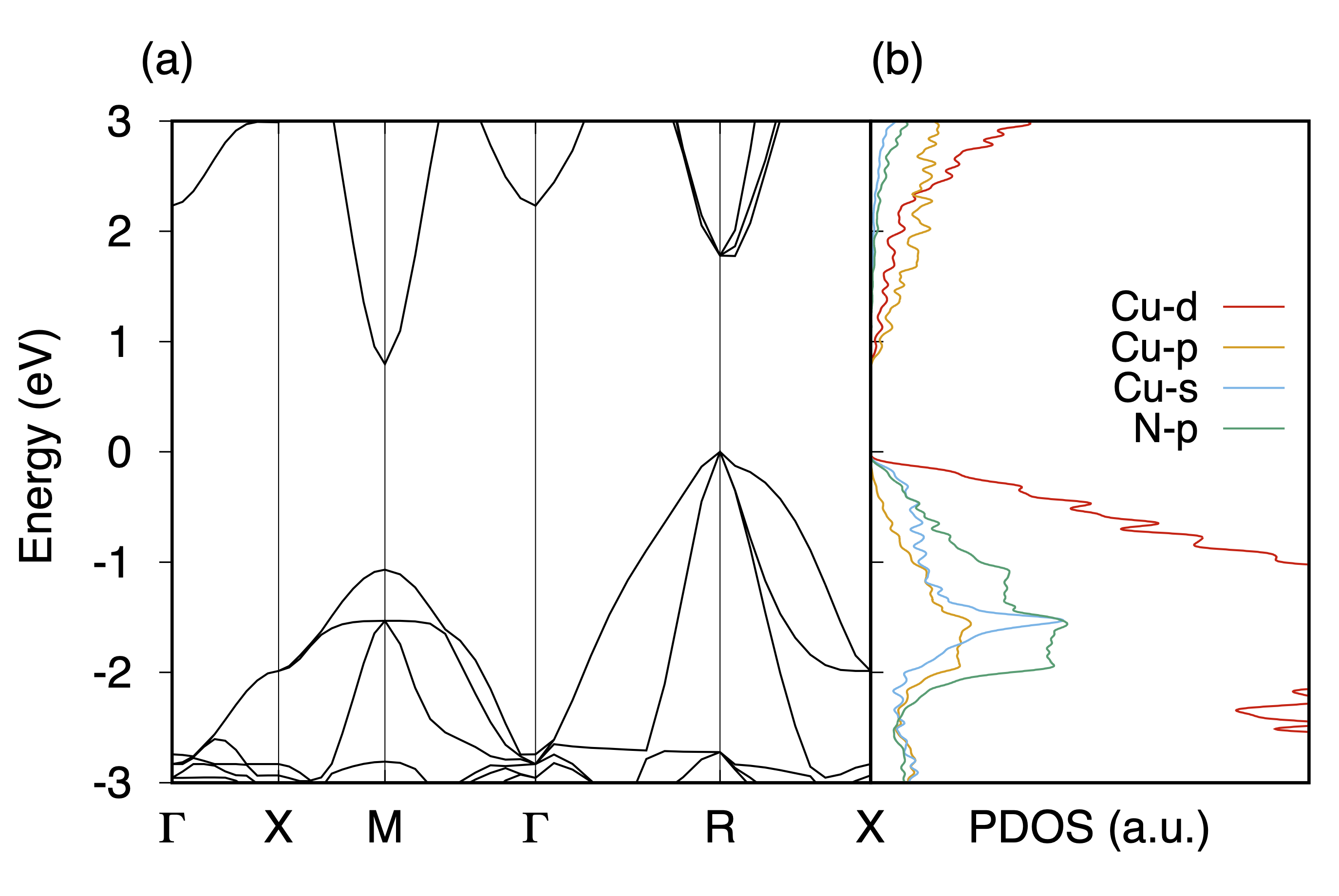}
\caption{(a) Band structure and (b) projected density of states (PDOS) for the primitive unit cell of Cu$_3$N computed with HSE. In (a), the band gap is indirect between the R and M points of the Brillouin zone. In (b), the PDOS is computed using $32 \times 32 \times 32$ \textbf{k}-points. The Fermi energy is set to the valence band maximum at R.}
\end{figure}

\begin{figure}[H]
\addcontentsline{toc}{figure}{Figure S2}
\centering
\includegraphics[width=0.7\linewidth]{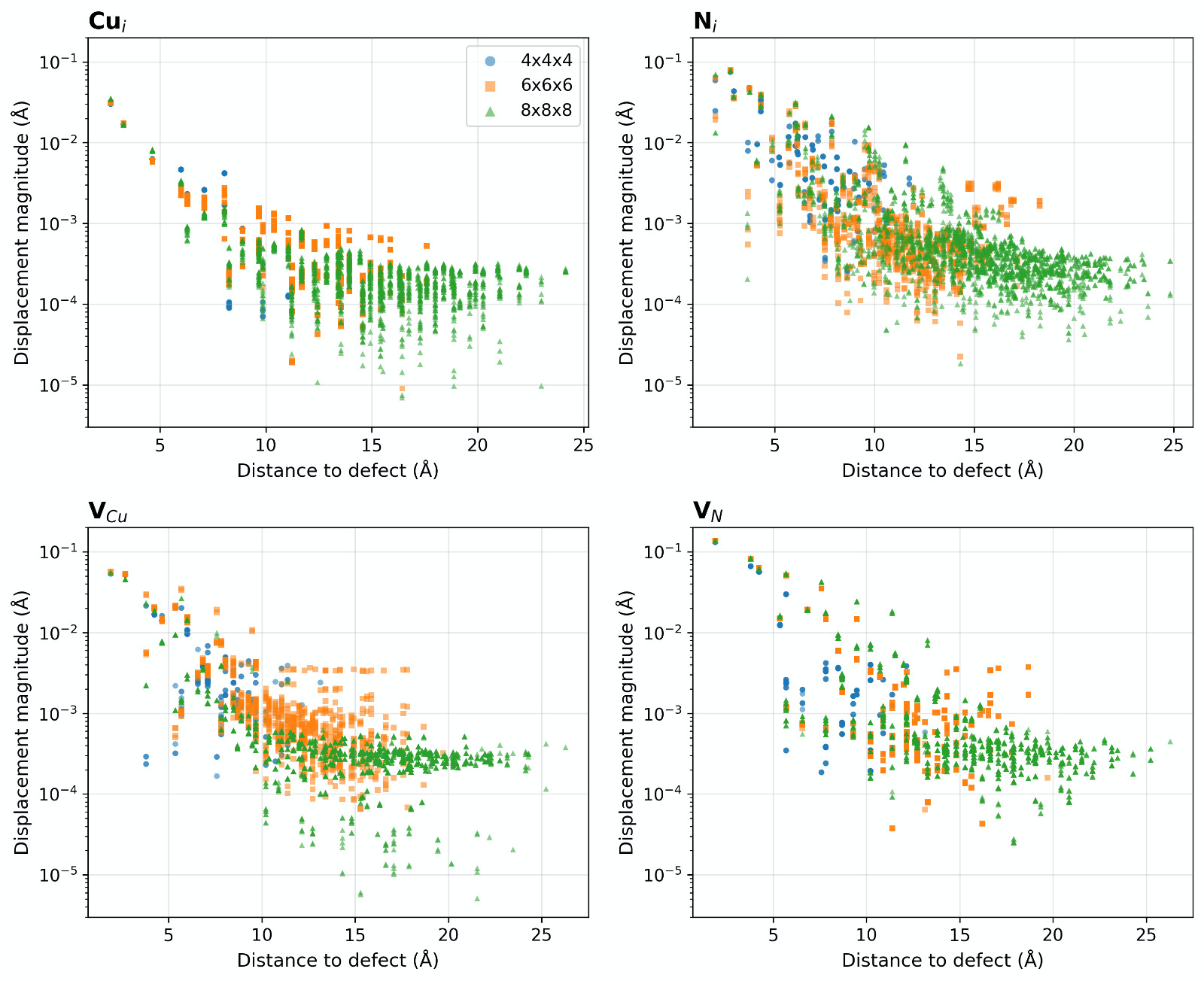}
\caption{Displacement magnitude of every atom vs. real-space distance to the defect site, log scale, for the three supercell sizes overlaid. Points beyond a given supercell size may reflect proximity to a periodic image of the defect rather than the true isolated-defect relaxation and should be read with that caveat. Vertical axis range identical across panels ($3 \times 10^{-6} - 2 \times 10^{-1}$ \AA{}).}
\end{figure}

\newpage

\begin{figure}
\addcontentsline{toc}{figure}{Figure S3}
\centering
\includegraphics[width=0.7\linewidth]{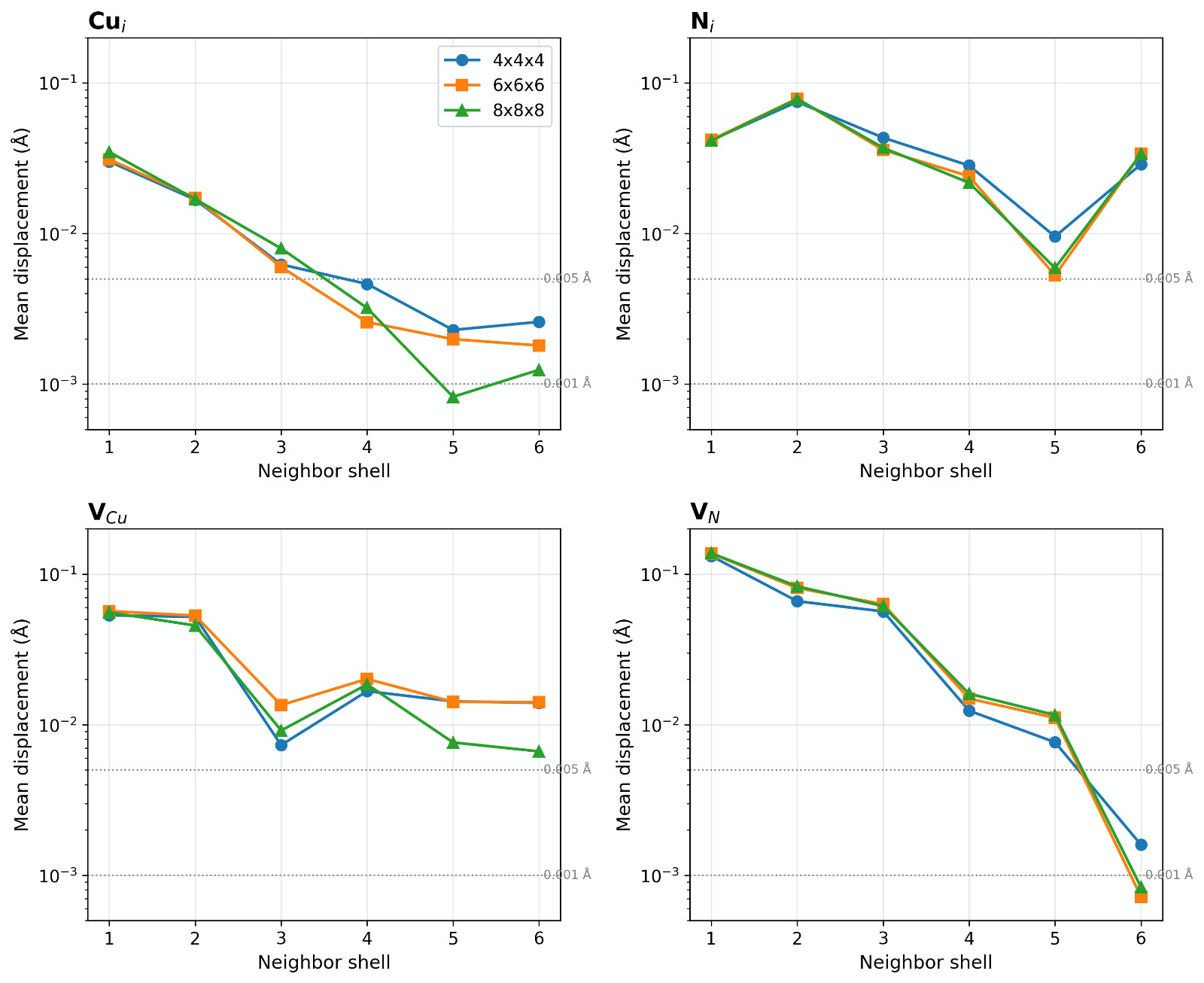}
\caption{Mean atomic displacement $\langle |\Delta \mathbf{r}| \rangle$ as a function of the neighbor-shell index $n$ for each neutral defect and supercell size. The plotted values are those listed in Table~S1, where the shell construction is described. Only the first six shells are shown, which lie within $7.1$~\AA{} of the defect. This keeps all atoms below half the side of the $4 \times 4 \times 4$ supercell ($L/2 \sim 7.6$~\AA{}), where the minimum-image convention ensures that each atom is closer to the defect than any of its periodic images. Within each panel, a given $n$ corresponds to the same coordination shell in the three supercells, as shown by the shell distances in Table~S1. Dotted gray lines mark 0.001 and 0.005~\AA{} for reference.}
\end{figure}

\newpage

\begin{figure}
\addcontentsline{toc}{figure}{Figure S4}
\centering
\includegraphics[width=0.8\linewidth]{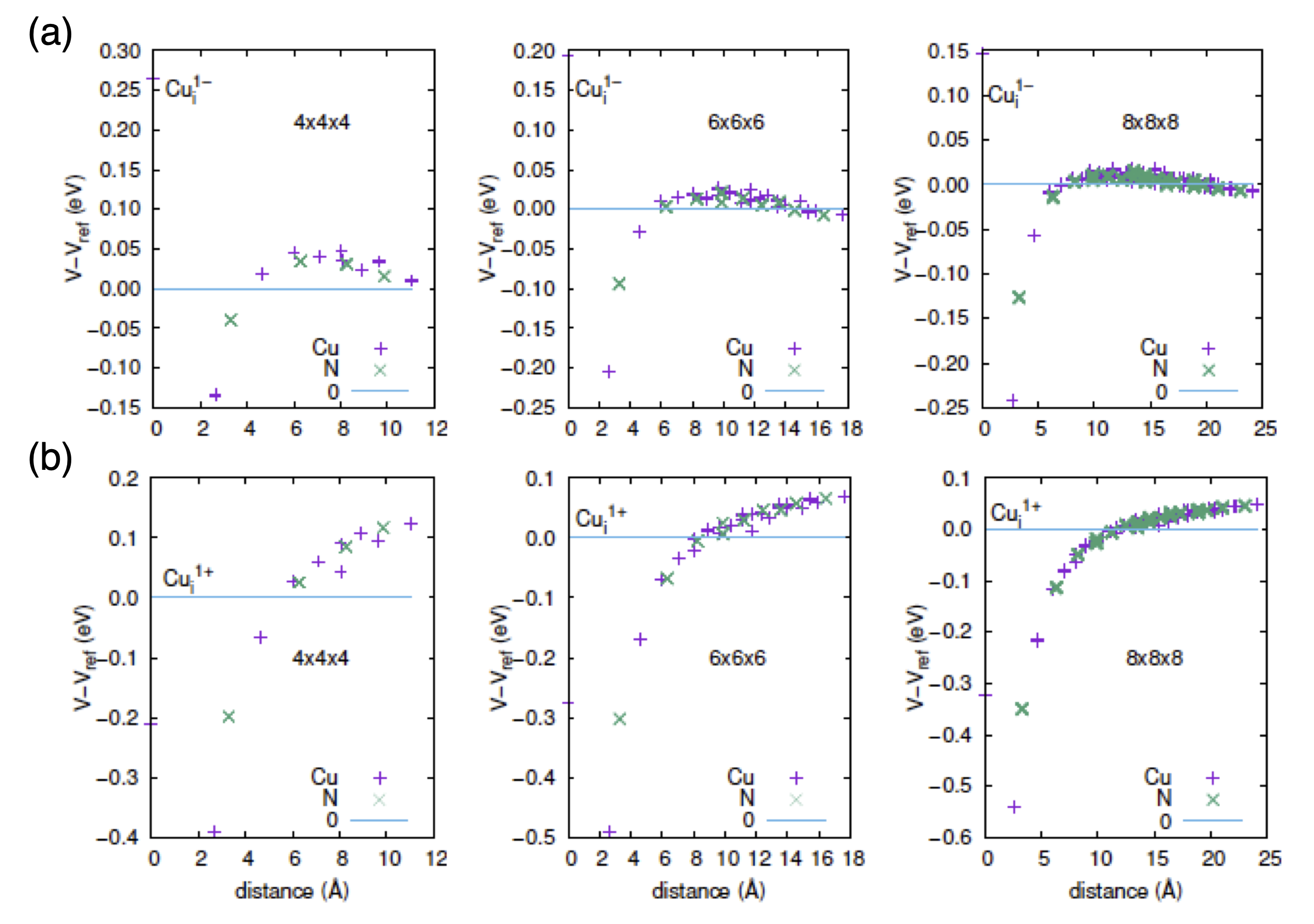}
\caption{Core level average potentials relative to the pristine material, as a function of the distance to (a) Cu$_{\textrm{i}}^{1-}$ and (b) Cu$_{\textrm{i}}^{1+}$. For each case, we show the results for $4 \times 4 \times 4$ (left), $6\times6\times6$ (center), and $8\times8\times8$ (right) supercell sizes. The horizontal line denotes the zero potential energy.}
\end{figure}

\begin{figure}
\addcontentsline{toc}{figure}{Figure S5}
\centering
\includegraphics[width=0.7\linewidth]{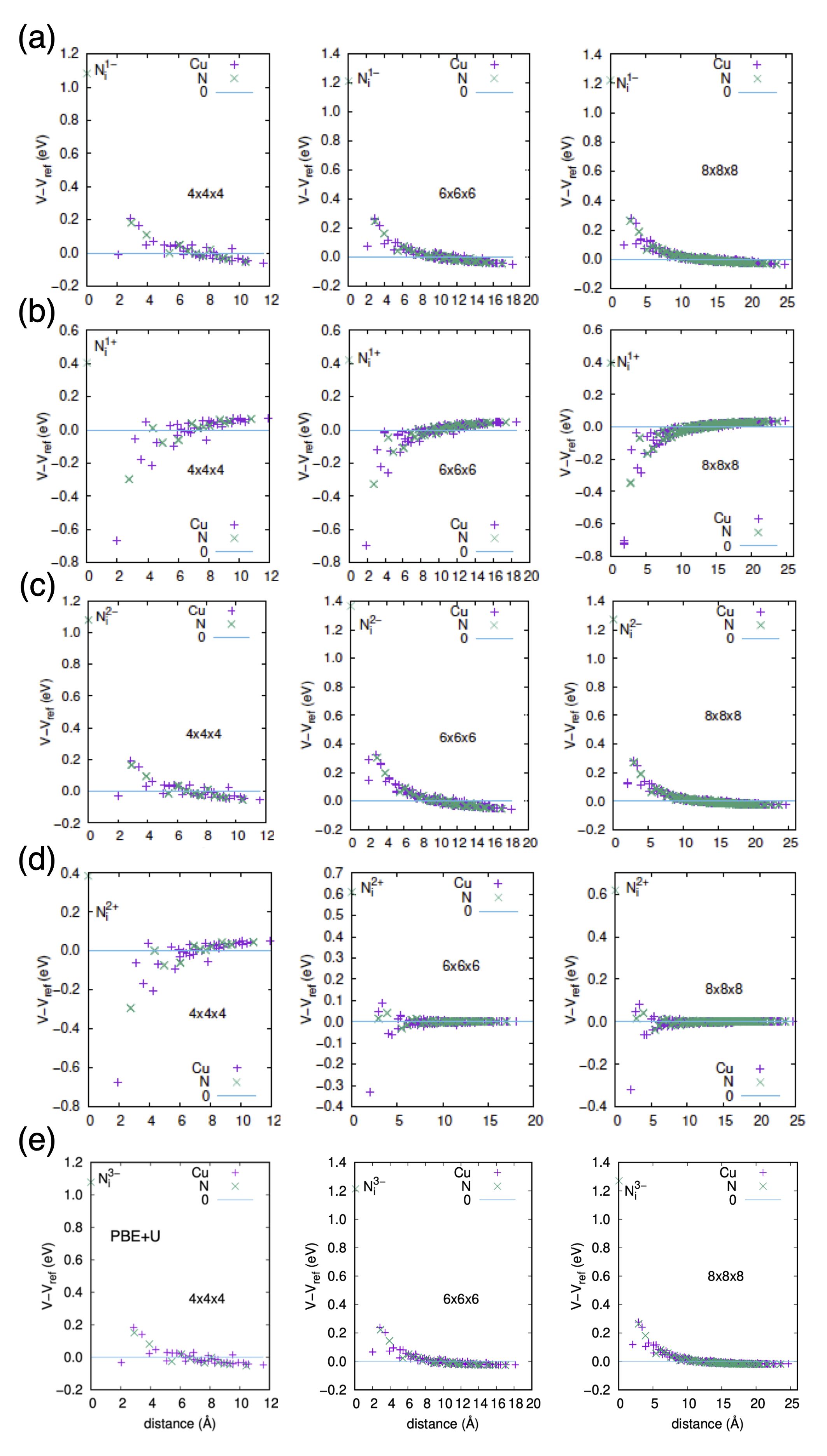}
\caption{Core level average potentials relative to the pristine material, as a function of the distance to (a) N$_{\textrm{i}}^{1-}$, (b) N$_{\textrm{i}}^{1+}$, (c) N$_{\textrm{i}}^{2-}$, (d) N$_{\textrm{i}}^{2+}$ and (e) N$_{\textrm{i}}^{3-}$. For each case, we show the results for $4 \times 4 \times 4$ (left), $6\times6\times6$ (center), and $8\times8\times8$ (right) supercell sizes. The horizontal line denotes the zero potential energy.}
\end{figure}

\begin{figure}
\addcontentsline{toc}{figure}{Figure S6}
\centering
\includegraphics[width=0.8\linewidth]{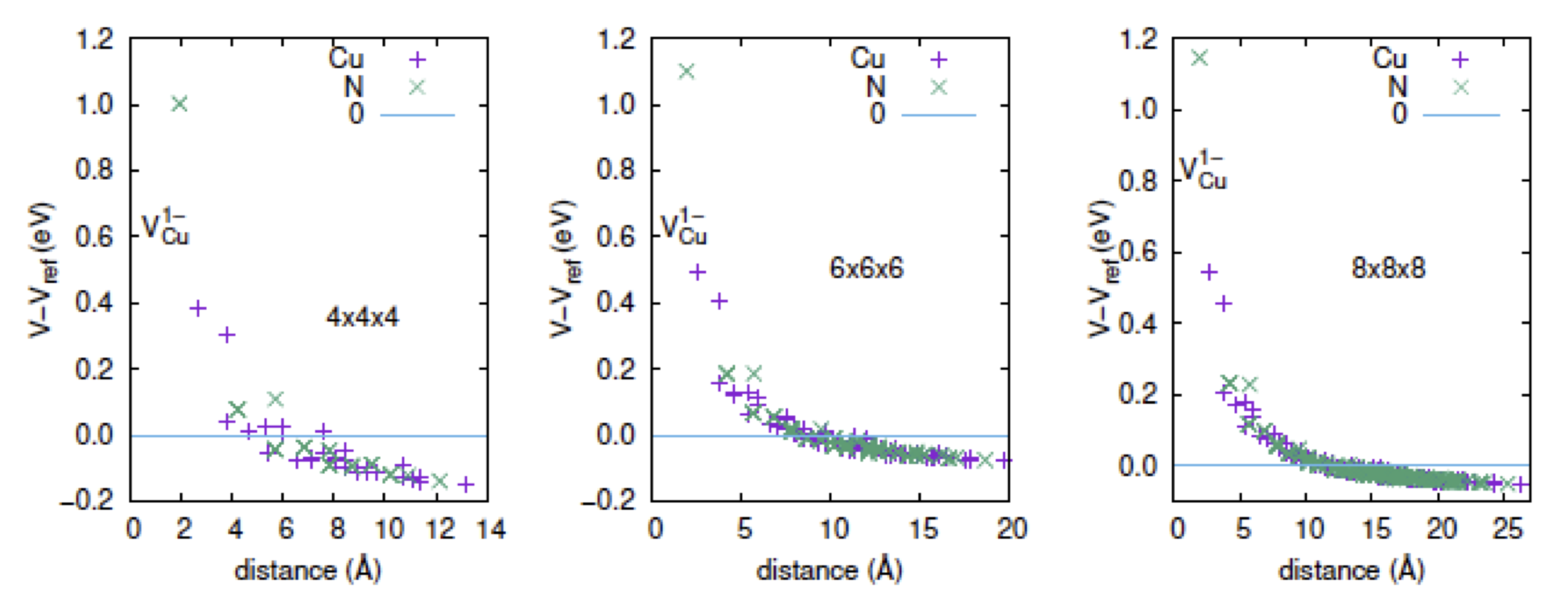}
\caption{Core level average potentials relative to the pristine material, as a function of the distance to V$_{\textrm{Cu}}^{1-}$, for $4 \times 4 \times 4$ (left), $6\times6\times6$ (center), and $8\times8\times8$ (right) supercell sizes. The horizontal line denotes the zero potential energy.}
\end{figure}

\begin{figure}
\addcontentsline{toc}{figure}{Figure S7}
\centering
\includegraphics[width=0.8\linewidth]{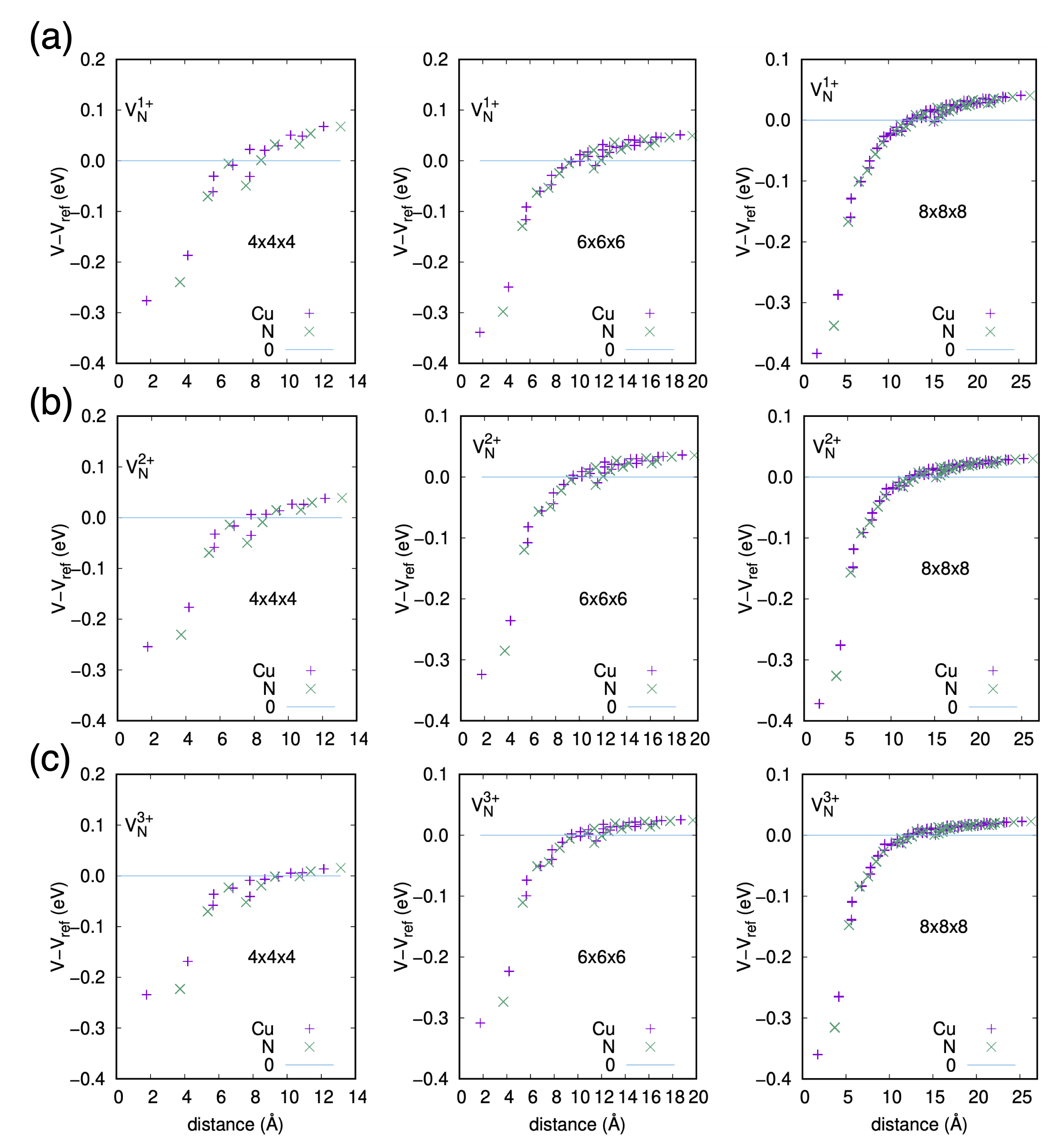}
\caption{Core level average potentials relative to the pristine material, as a function
of the distance to (a) V$_{\textrm{N}}^{1+}$, (b) V$_{\textrm{N}}^{2+}$ and (c) V$_{\textrm{N}}^{3+}$. For each case, we show the results for $4 \times 4 \times 4$ (left), $6\times6\times6$ (center), and $8\times8\times8$ (right) supercell sizes. The horizontal line denotes the zero potential energy.}
\end{figure}

\begin{figure*}
\addcontentsline{toc}{figure}{Figure S8}
\centering
\includegraphics[width=\linewidth]{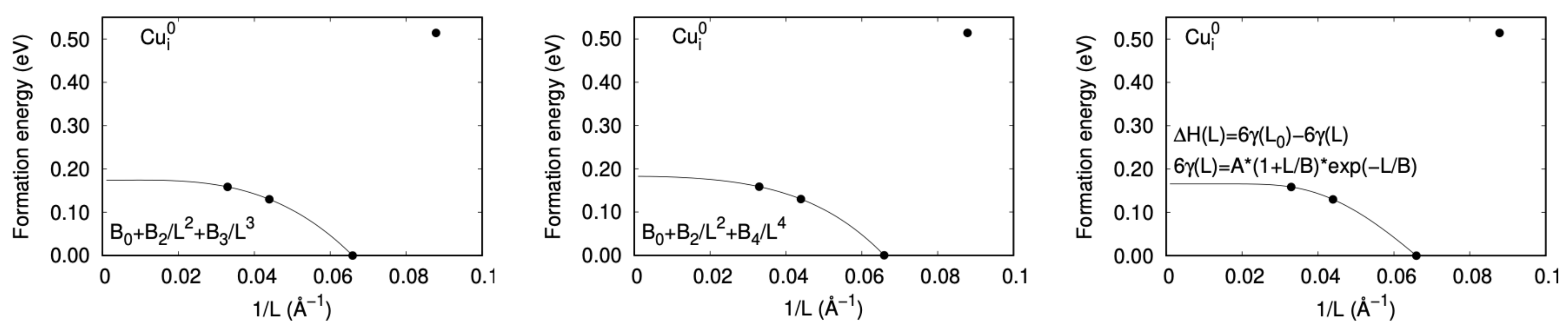}
\caption{Fittings for the formation energy of Cu$_{\textrm{i}}^{0}$ using a cubic polynomial (left), quartic polynomial (center) and the impurity model (right). The size corrections derived are 0.174 eV (left), 0.182 eV (center), and 0.166 eV (right). The outlier at (0.09,0.51) was not included in the fitting.}
\end{figure*}

\begin{figure*}
\addcontentsline{toc}{figure}{Figure S9}
\centering
\includegraphics[width=0.7\linewidth]{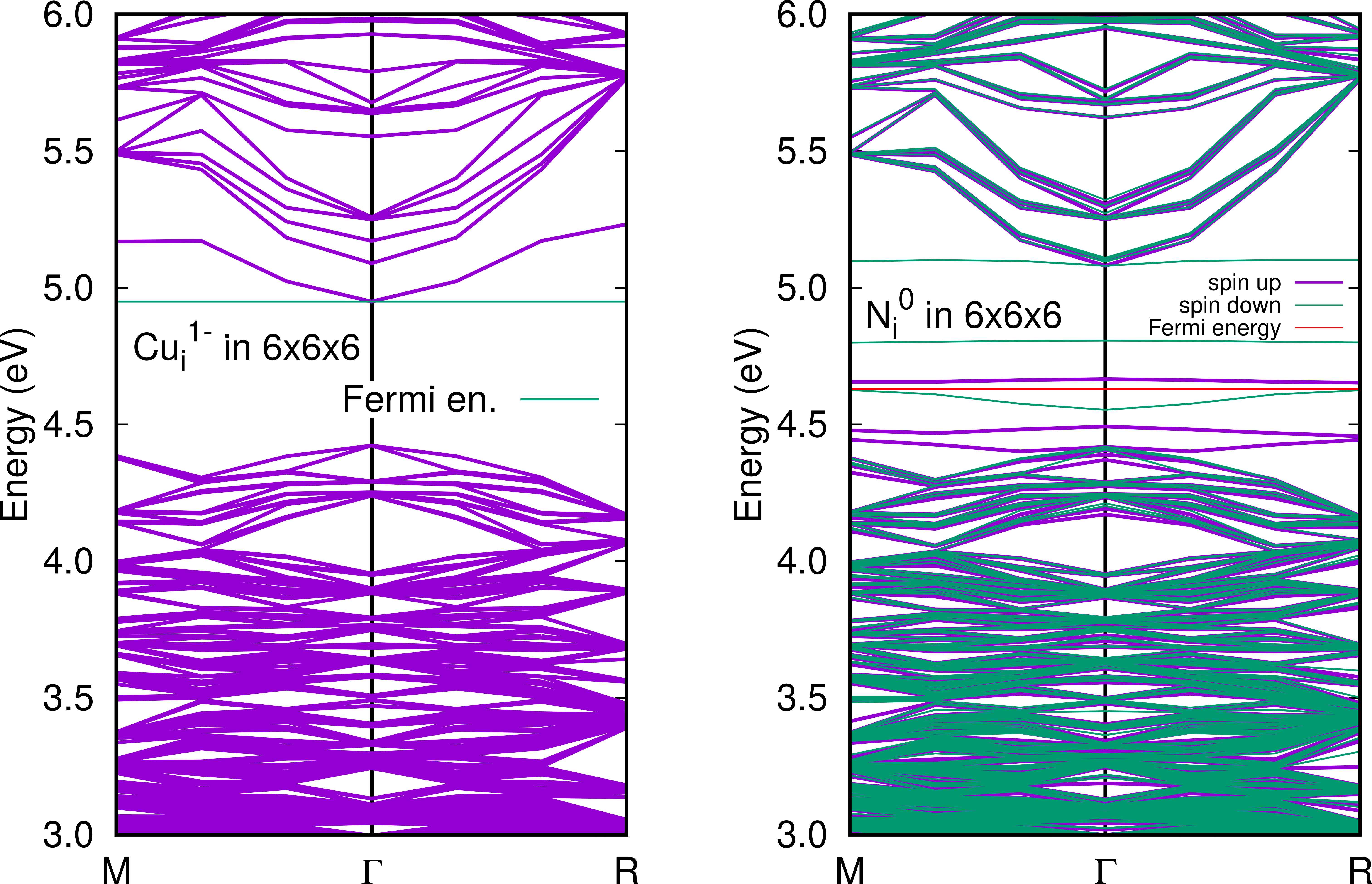}
\caption{Band structure of Cu$_\textrm{i}^{1-}$ (left panel) and N$_\textrm{i}^{0}$ (right panel) within a $6 \times 6 \times 6$ supercell. For Cu$_\textrm{i}^{1-}$, the band structure is computed without spin polarization and with only a few \textbf{k}-points to avoid the high computational cost. For N$_\textrm{i}^{0}$, the bandwidth of the in-gap state is strongly reduced, reversing its energy order with respect to the $4\times4\times4$ supercell.
}
\end{figure*}

\begin{figure*}
\addcontentsline{toc}{figure}{Figure S10}
 \includegraphics[width=\linewidth]{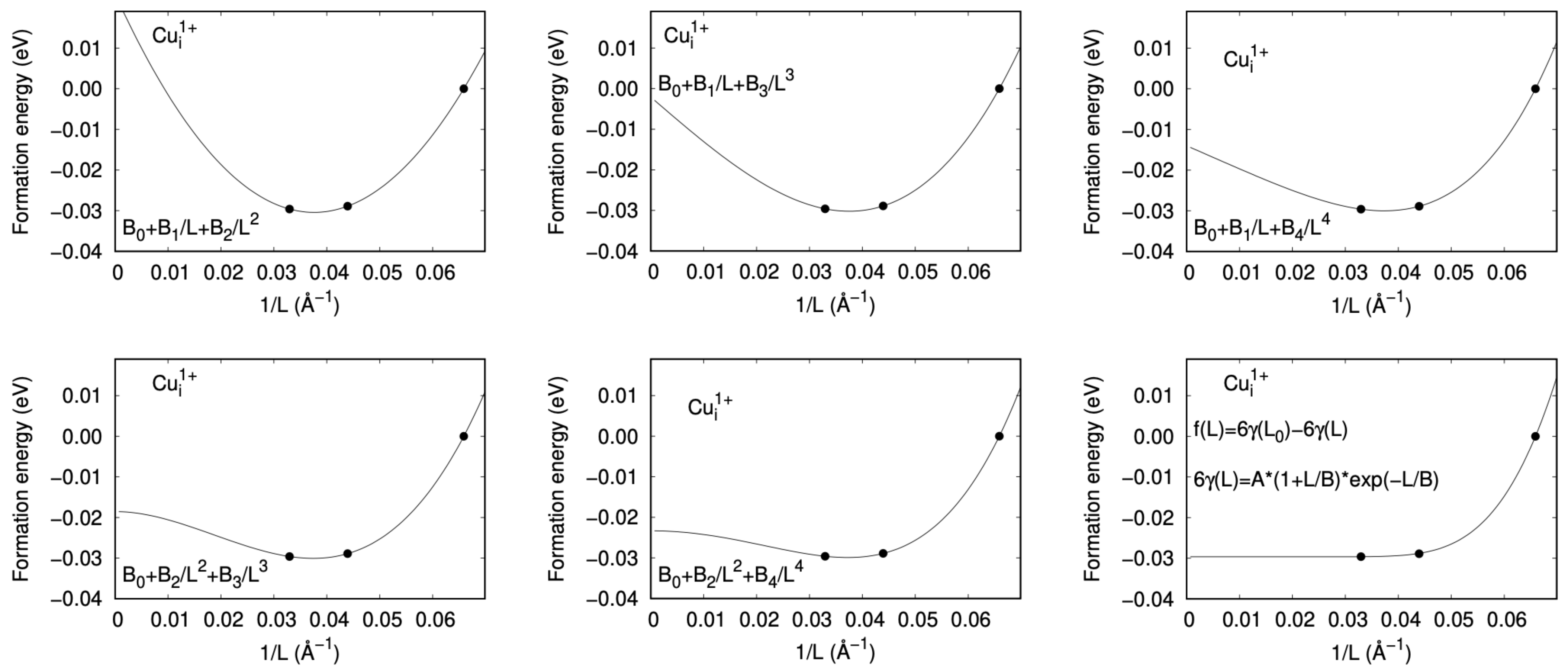}
\caption{Finite-size extrapolation of the formation energy of Cu$_{\mathrm{i}}^{1+}$ using polynomial and hydrogen-like impurity-model fits. The size  corrections obtained are, from left to right, top to bottom: 0.023 eV, -0.002 eV, -0.014 eV, -0.019 eV, -0.023 eV, and -0.030 eV. The adopted correction is the intermediate value of 0.0~eV, obtained from the Makov--Payne-like expression $B_0+B_1/L+B_3/L^3$. The hydrogen-like impurity model yields a correction of -0.030~eV, but this model is not physically justified for Cu$_{\mathrm{i}}^{1+}$ because the in-gap defect band is unoccupied.}
\end{figure*}

\begin{figure*}
\addcontentsline{toc}{figure}{Figure S11}
\centering
\includegraphics[width=0.8\linewidth]{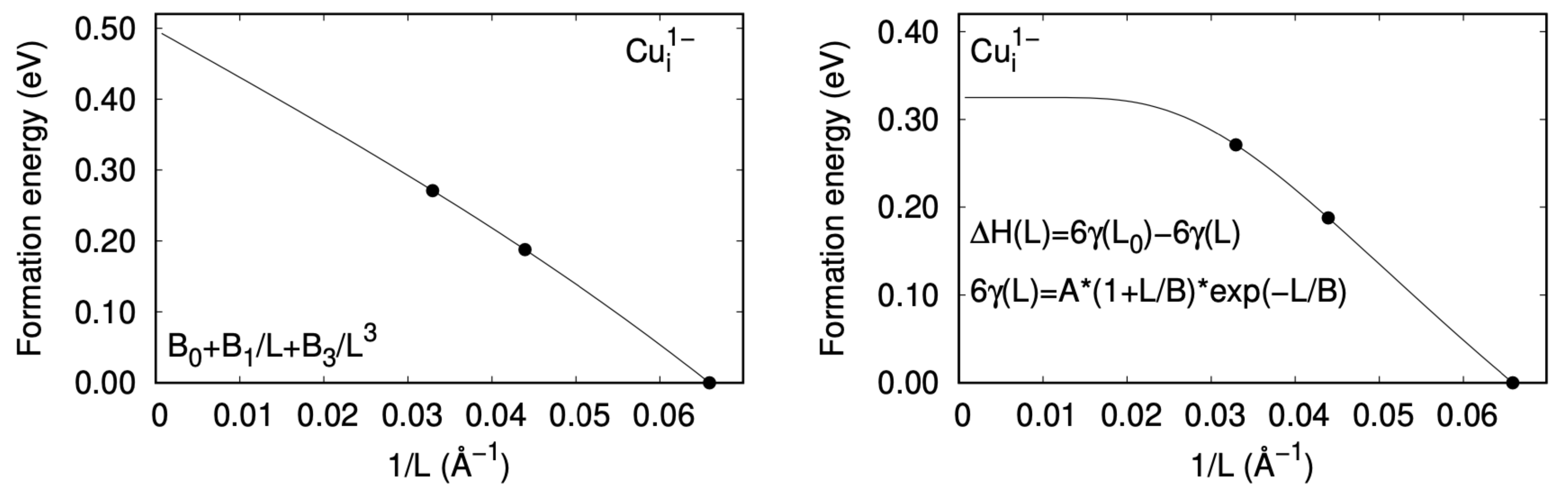}
\caption{Fittings for the formation energy of Cu$_{\textrm{i}}^{1-}$ using a cubic polynomial (left) and the impurity model (right). The size corrections derived are 0.498~eV (left) and 0.325~eV (right).}
\end{figure*}

\begin{figure*}
\addcontentsline{toc}{figure}{Figure S12}
\centering
\includegraphics[width=0.8\linewidth]{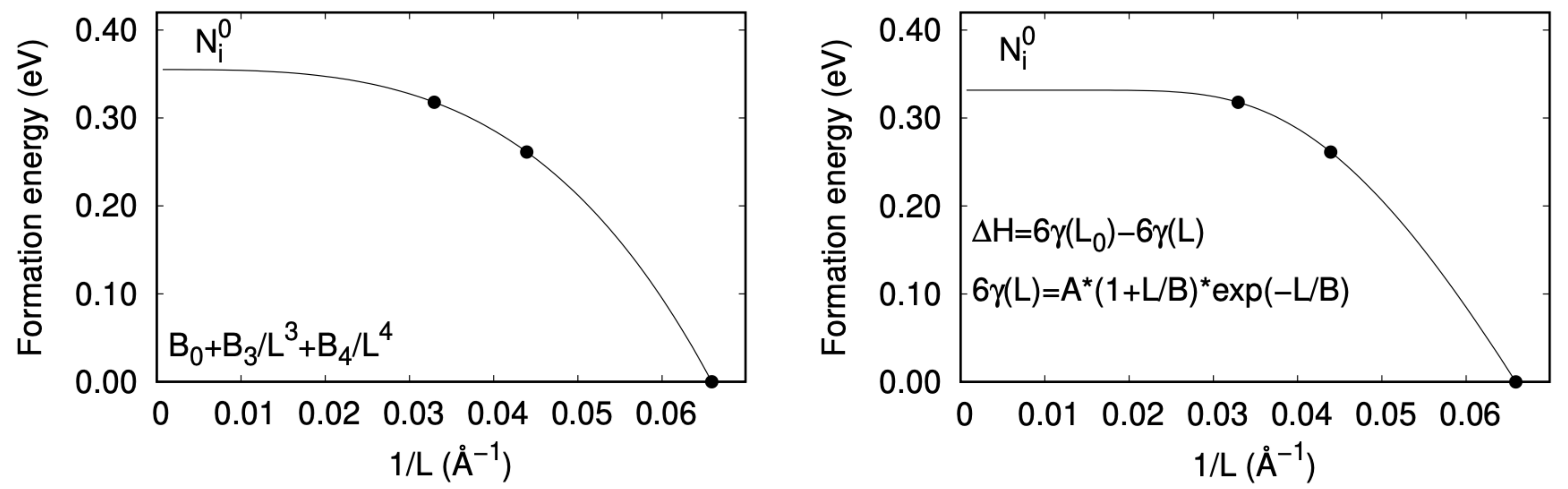}
\caption{Fittings for the formation energy of N$_{\textrm{i}}^{0}$ using a quartic polynomial (left) and the impurity model (right). The size corrections derived are 0.355 eV (left) and 0.332 eV (right).}
\end{figure*}

\begin{figure*}
\addcontentsline{toc}{figure}{Figure S13}
\centering
\includegraphics[width=\linewidth]{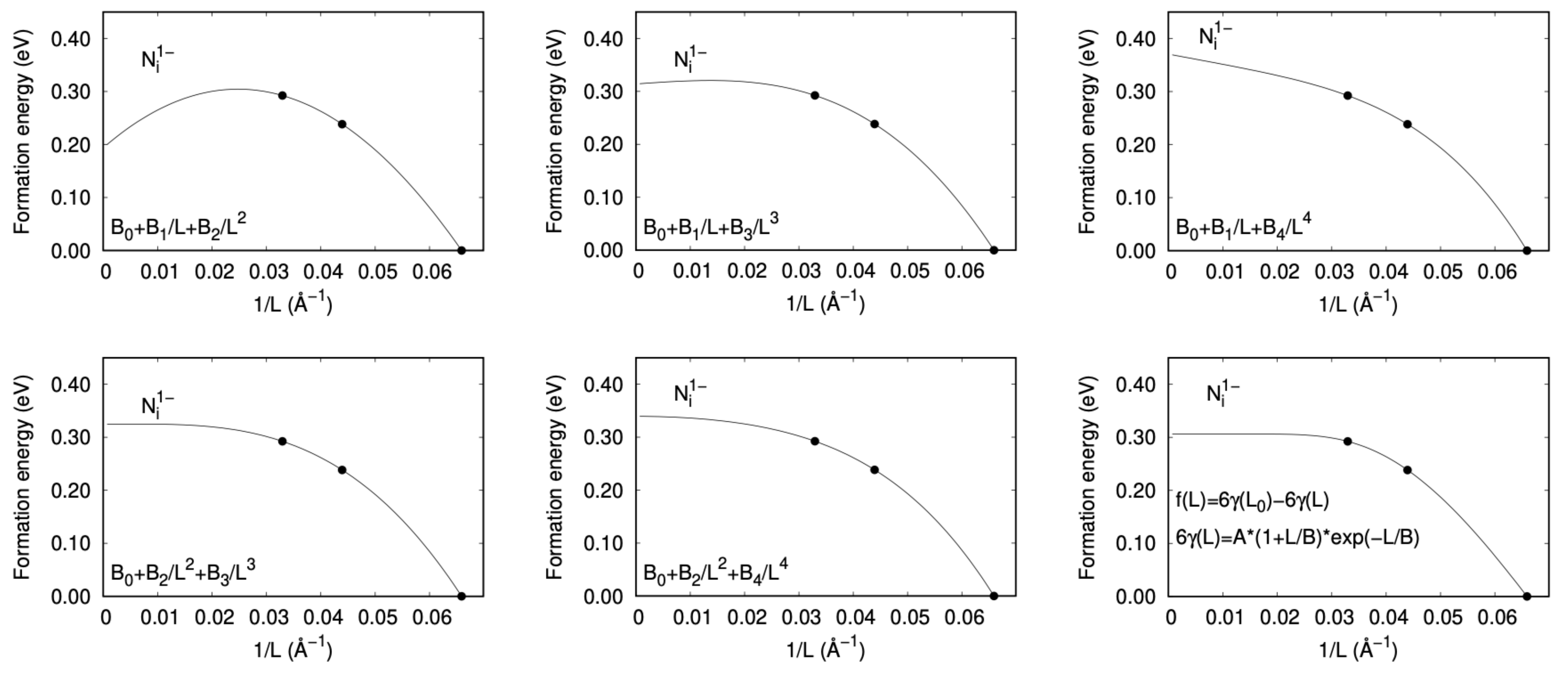}
\caption{Fittings for the formation energy of N$_{\textrm{i}}^{1-}$ using different polynomials and the impurity model. The size corrections derived are, from left to right, top to bottom: 0.194 eV, 0.314 eV, 0.370 eV, 0.324 eV, 0.339 eV, and 0.306 eV.}
\end{figure*}

\begin{figure*}
\addcontentsline{toc}{figure}{Figure S14}
\centering
\includegraphics[width=\linewidth]{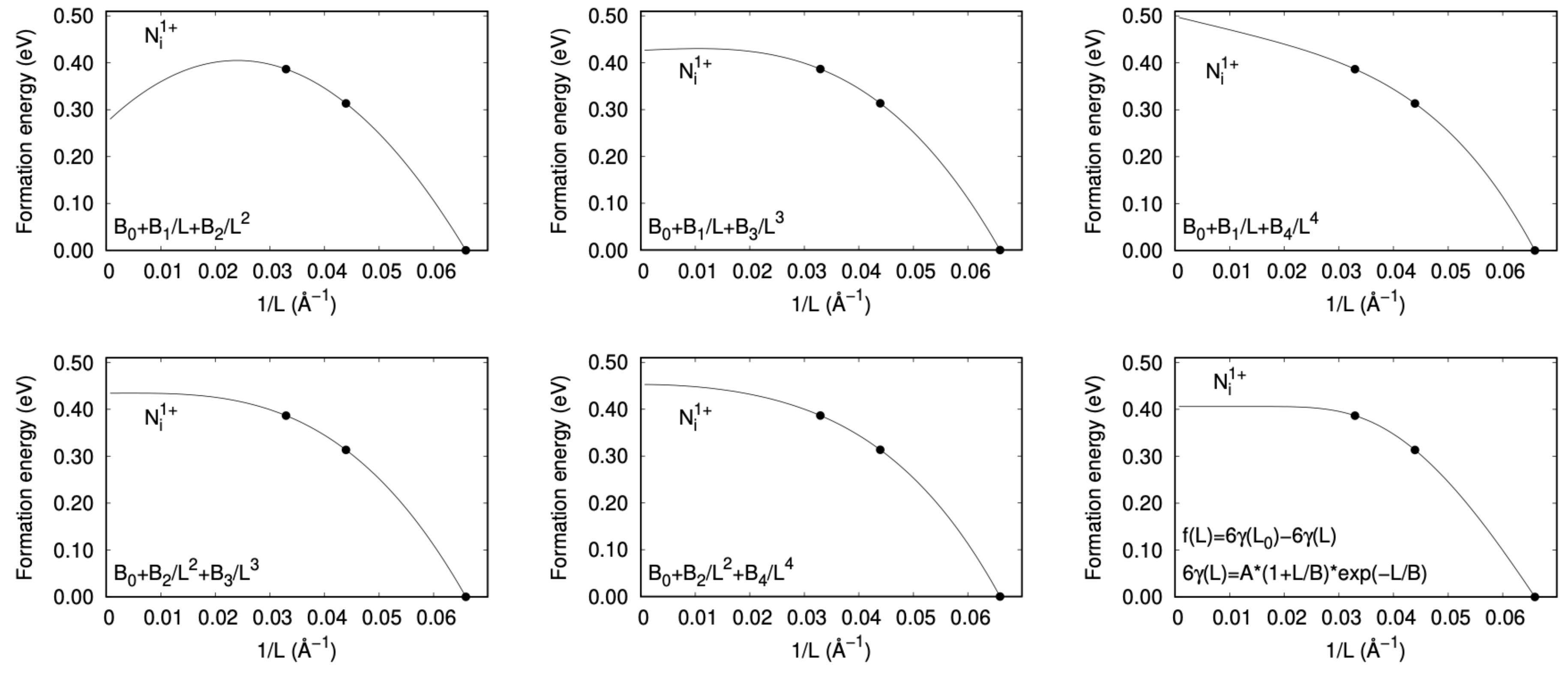}
\caption{Fittings for the formation energy of N$_{\textrm{i}}^{1+}$ using different polynomials and the impurity model. The size corrections derived are, from left to right, top to bottom: 0.272~eV, 0.426~eV, 0.498~eV, 0.434~eV, 0.453~eV, and 0.406~eV.}
\end{figure*}

\begin{figure*}
\addcontentsline{toc}{figure}{Figure S15}
\centering
\includegraphics[width=\linewidth]{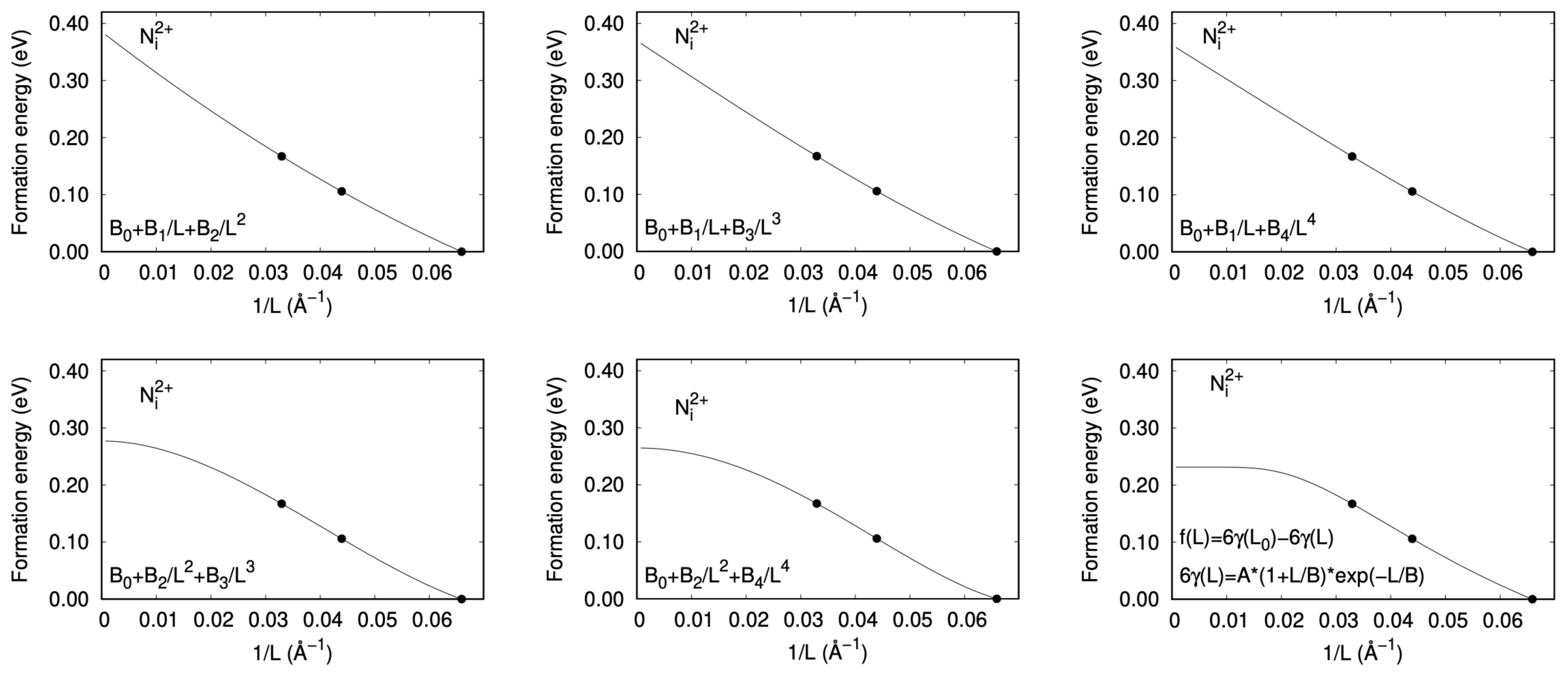}
\caption{Fittings for the formation energy of N$_{\textrm{i}}^{2+}$ using different polynomials and the impurity model. The size corrections derived are, from left to right, top to bottom: 0.385~eV, 0.370~eV , 0.362~eV , 0.277~eV, 0.265~eV, and 0.231~eV.}
\end{figure*}

\begin{figure*}
\addcontentsline{toc}{figure}{Figure S16}
\centering
\includegraphics[width=\linewidth]{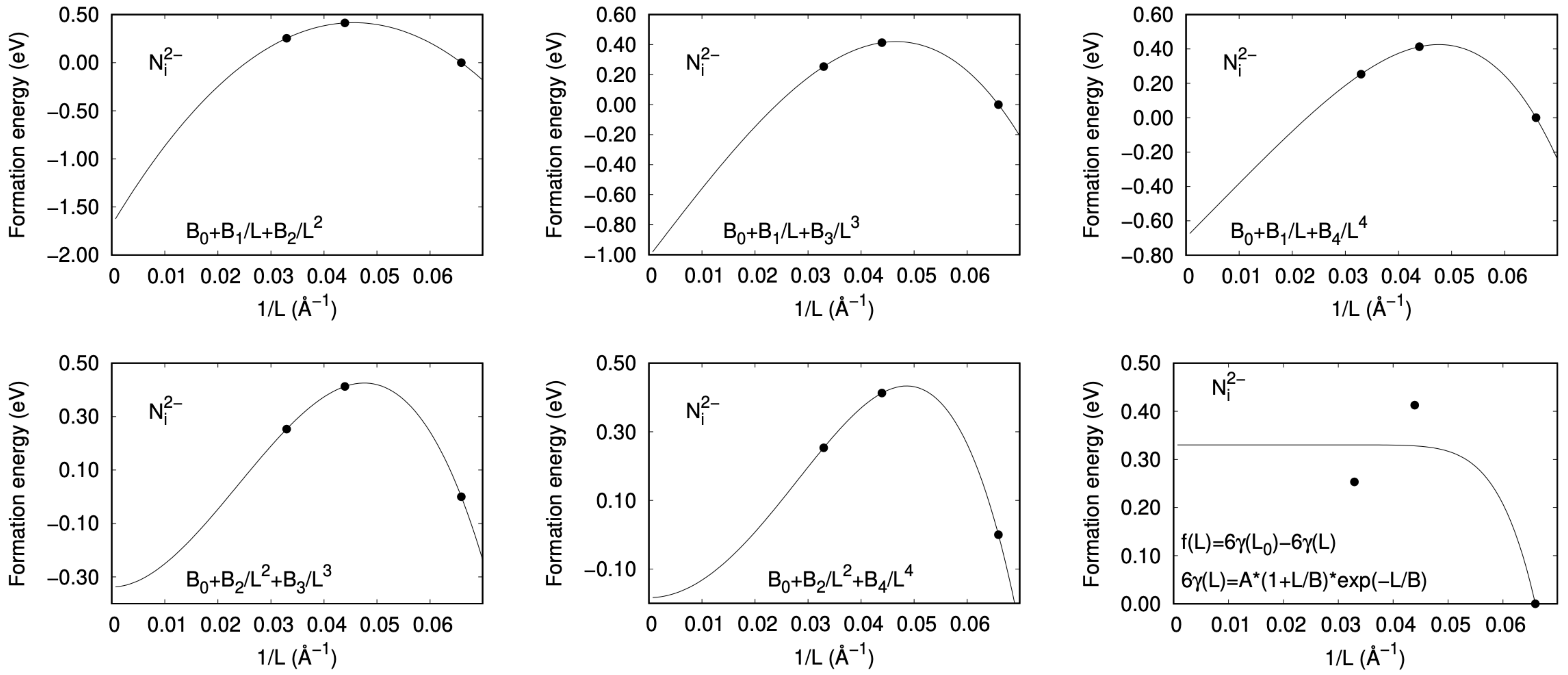}
\caption{Fittings for the formation energy of N$_{\textrm{i}}^{2-}$ using different polynomials and the impurity model. None of these models can be considered satisfactory.}
\end{figure*}

\begin{figure*}
\addcontentsline{toc}{figure}{Figure S17}
\centering
\includegraphics[width=\linewidth]{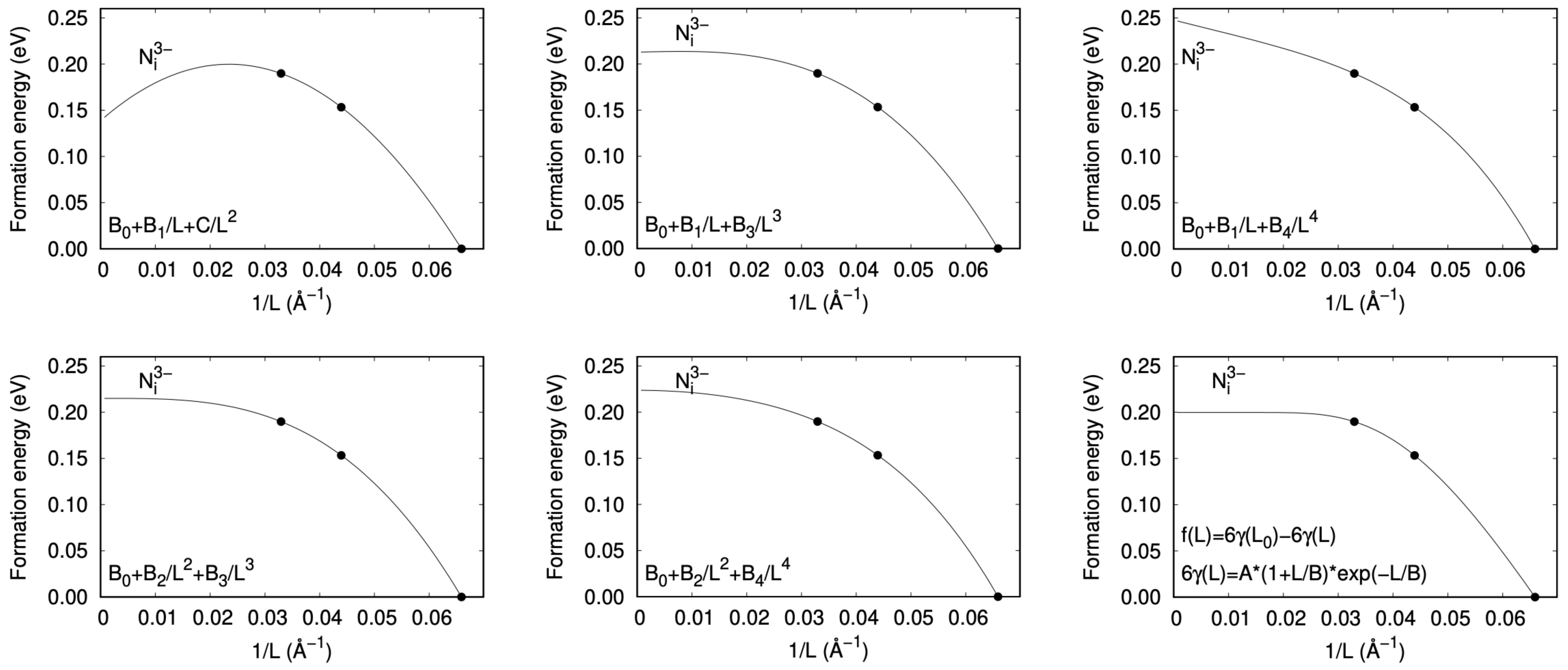}
\caption{Finite-size extrapolation of the formation energy of N$_{\mathrm{i}}^{3-}$ using polynomial and hydrogen-like impurity-model fits. The size corrections are, from left to right, and top to bottom:0.139 eV, 0.213 eV, 0.247 eV, 0.215 eV, 0.224~eV, and 0.200 eV.}
\end{figure*}

\begin{figure*}
\addcontentsline{toc}{figure}{Figure S18}
\centering
\includegraphics[width=\linewidth]{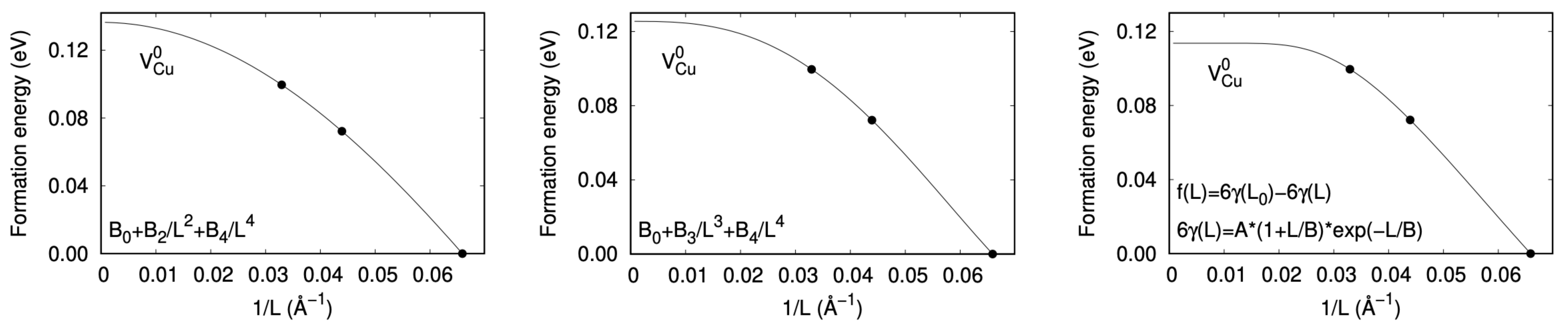}
\caption{Fittings for the formation energy of V$_{\textrm{Cu}}^{0}$ using a quartic polynomial without linear term (left), a quartic polynomial with linear term (center) and the impurity model (right). The size corrections derived are 0.136 eV (left), 0.125 eV (center) and 0.114 eV (right).}
\end{figure*}

\begin{figure*}
\addcontentsline{toc}{figure}{Figure S19}
\centering
\includegraphics[width=\linewidth]{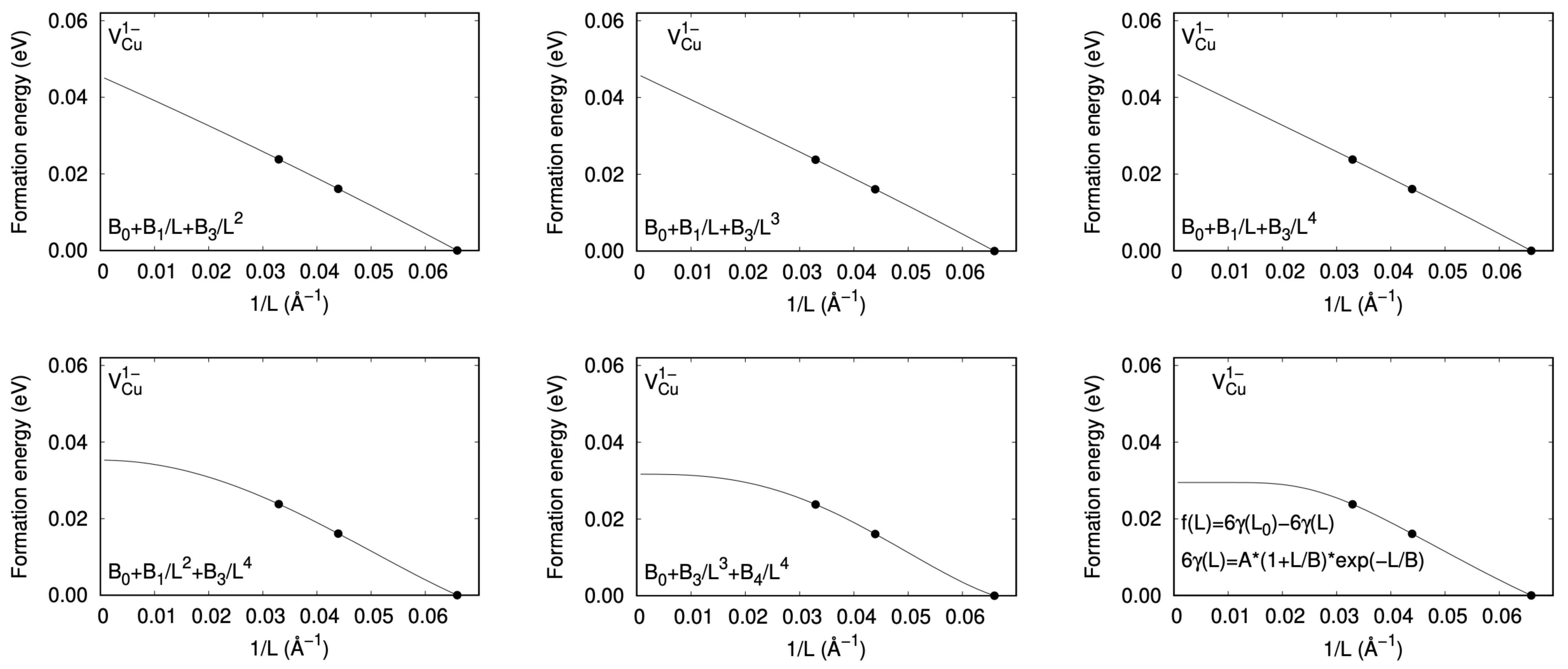}
\caption{Fittings for the formation energy of V$_\textrm{Cu}^{1-}$ using different polynomials and the impurity model. The size corrections derived are, from left to right, top to bottom: 0.045 eV, 0.046 eV, 0.046 eV, 0.035 eV, 0.032 eV and 0.029.}
\end{figure*}

\begin{figure*}
\addcontentsline{toc}{figure}{Figure S20}
\centering
\includegraphics[width=0.7\linewidth]{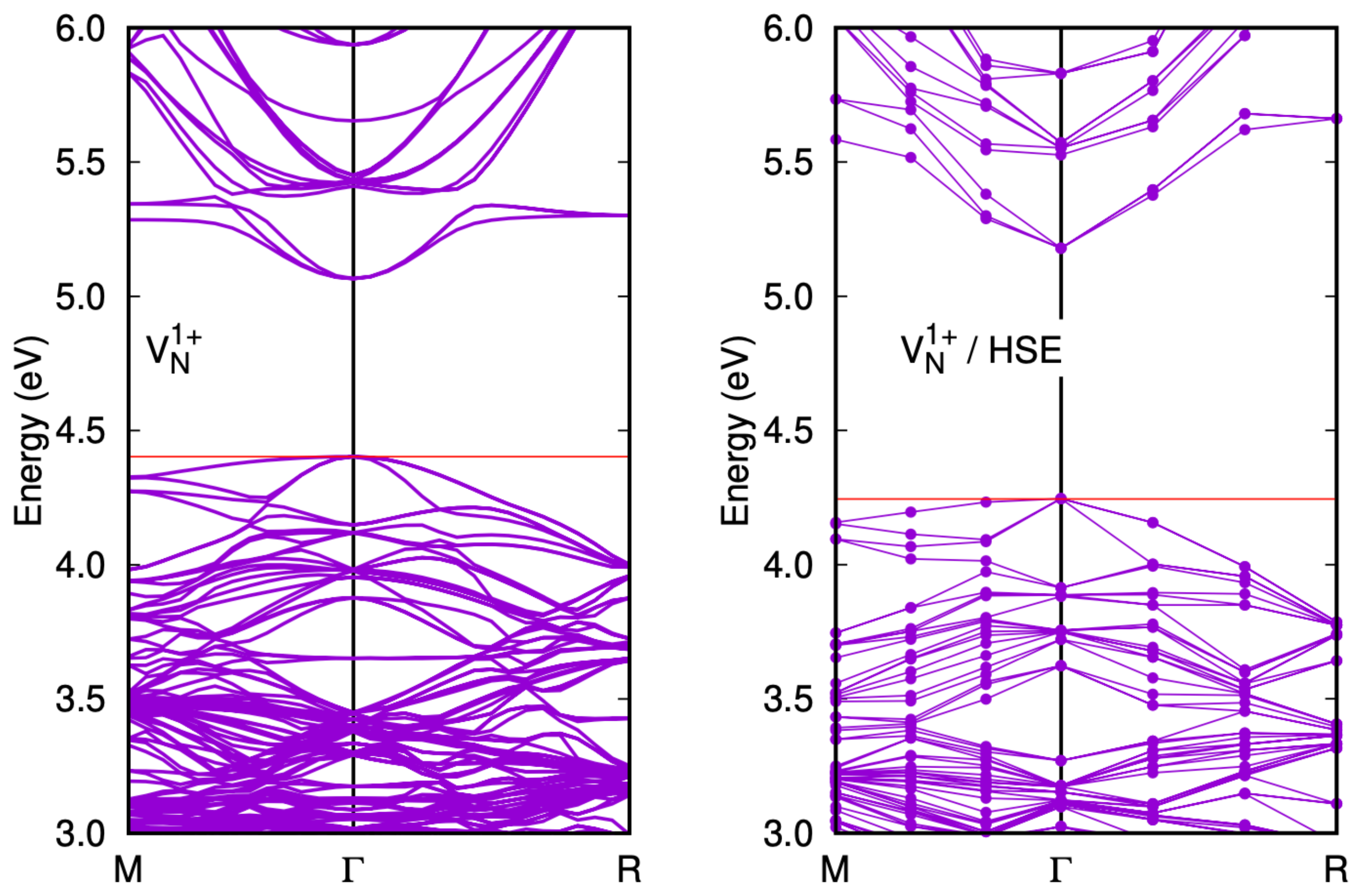}
\caption{Band structure of V$_\textrm{N}^{1+}$ with PBE+U (left panel) and HSE (right panel) in a $4\times 4\times 4$ supercell and without spin polarization. PBE+U and HSE are computed with 30 and 7 \textbf{k}-points between $M$ and $R$, respectively.}
\end{figure*}

\begin{figure*}
\addcontentsline{toc}{figure}{Figure S21}
\centering
\includegraphics[width=\linewidth]{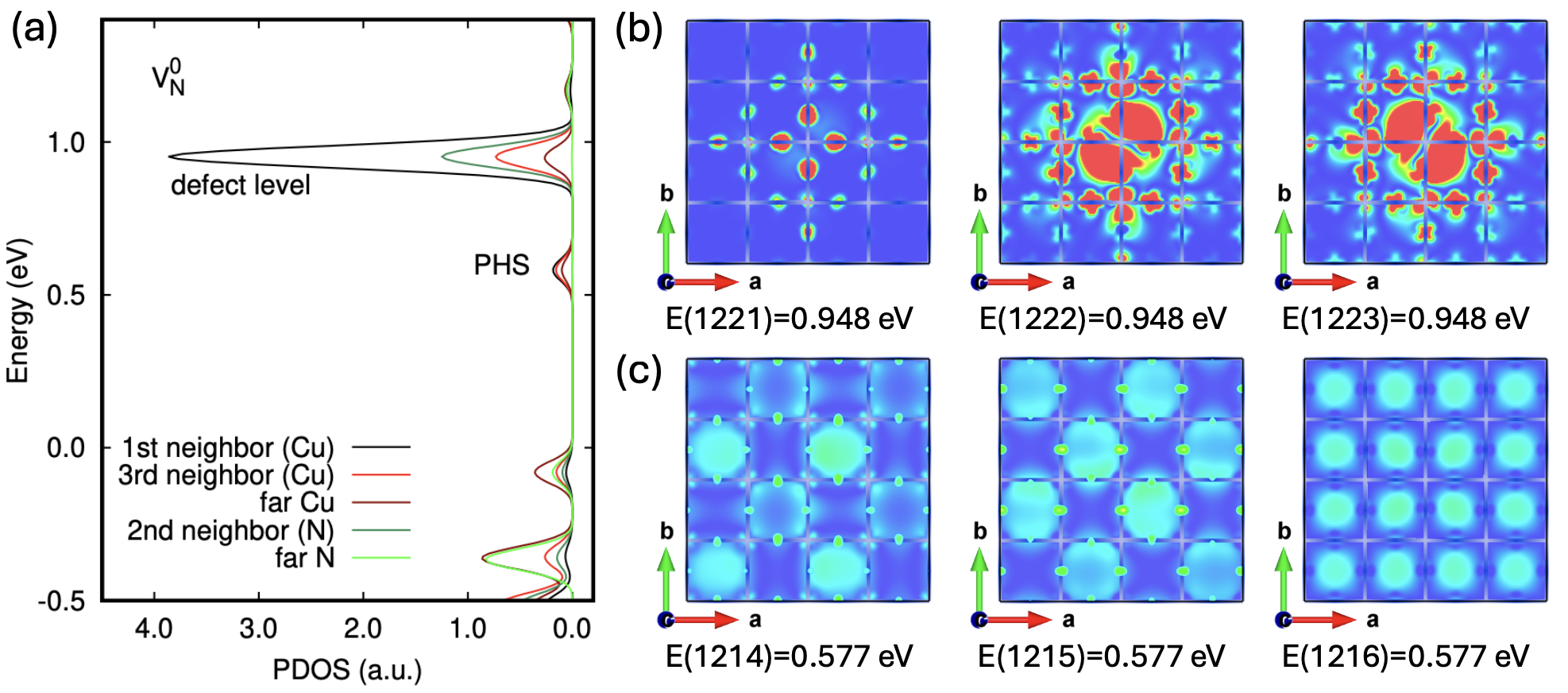}
\caption{(a) $\Gamma$-point partial density of states of \defeq{V}{N}{0} ($4\times4\times4$ supercell, PBE+U) projected onto individual atoms: first- and third-neighbor Cu, second-neighbor N, and far Cu and N. Energies relative to the pristine VBM. (b, c) Partial charge densities of the three degenerate states at 0.95~eV (b, defect states; slice at $z=0.5$, through the vacancy) and 0.58~eV (c, perturbed host state (PHS); slice at $z=0.375$, through the cavity centers to include the maximum density). Red: high density; blue: low density; color scale normalized per panel.}
\end{figure*}

\begin{figure*}
\addcontentsline{toc}{figure}{Figure S22}
\centering
\includegraphics[width=\linewidth]{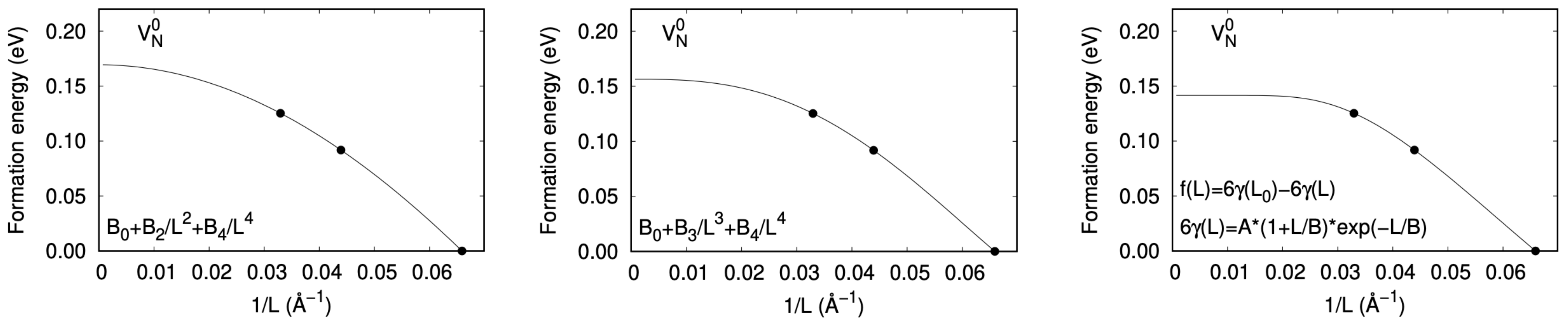}
\caption{Fittings for the formation energy of V$_\textrm{N}^{0}$ using a quartic polynomial without cubic term (left), quartic polynomial with cubic term (center) and the impurity model (right). The size corrections derived are 0.169~eV (left), 0.156~eV (center) and 0.142~eV (right).}
\end{figure*}

\begin{figure*}
\addcontentsline{toc}{figure}{Figure S23}
\centering
\includegraphics[width=\linewidth]{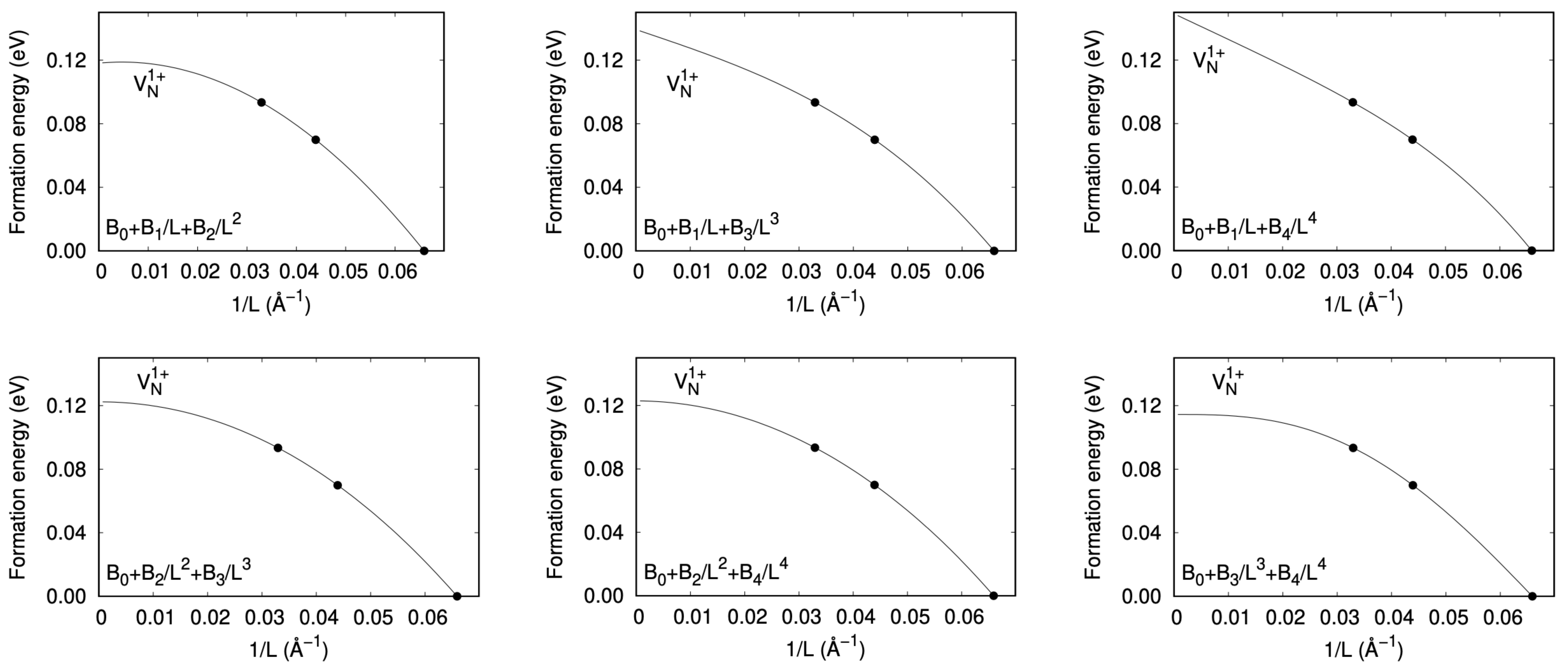}
\caption{Fittings for the formation energy of V$_\textrm{N}^{1+}$ using different polynomials.  The size corrections are, from left to right, and top to bottom: 0.118 eV, 0.139 eV, 0.149 eV, 0.122~eV, 0.123~eV, and 0.114~eV.}
\end{figure*}

\begin{figure*}
\addcontentsline{toc}{figure}{Figure S24}
\centering
\includegraphics[width=\linewidth]{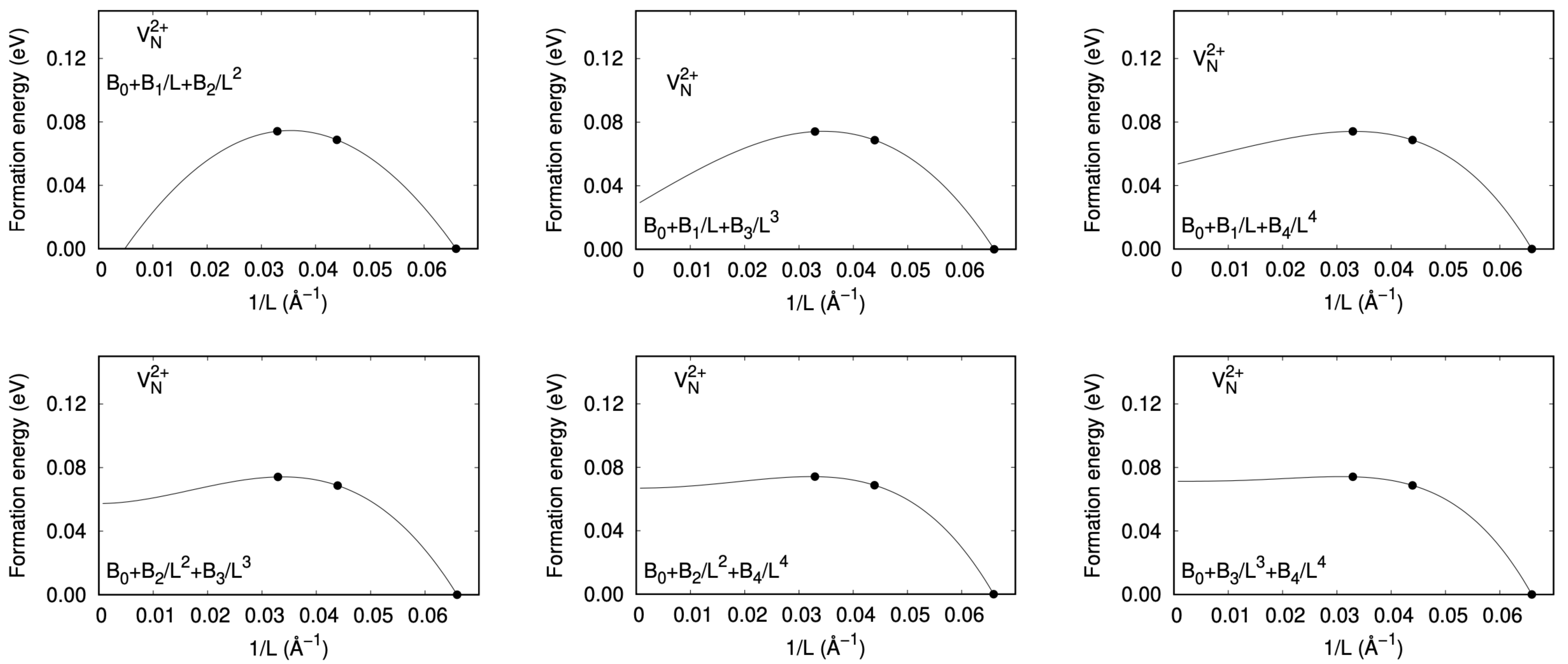}
\caption{Fittings for the formation energy of V$_\textrm{N}^{2+}$ using different polynomials.  The size corrections are, from left to right, and top to bottom:  -0.025~eV, 0.028~eV, 0.053~eV, 0.057~eV, 0.067~eV, and 0.071~eV. The Makov-Payne correction (top-center) gives a correction of 0.028~eV, but $B_{1}>0$ is in contradiction with the model. Instead, we choose as correction the difference between $8\times 8\times 8$ and $4\times 4\times 4$ supercells, i.e., 0.074~eV.}
\end{figure*}

\begin{figure*}
\addcontentsline{toc}{figure}{Figure S25}
\centering
\includegraphics[width=\linewidth]{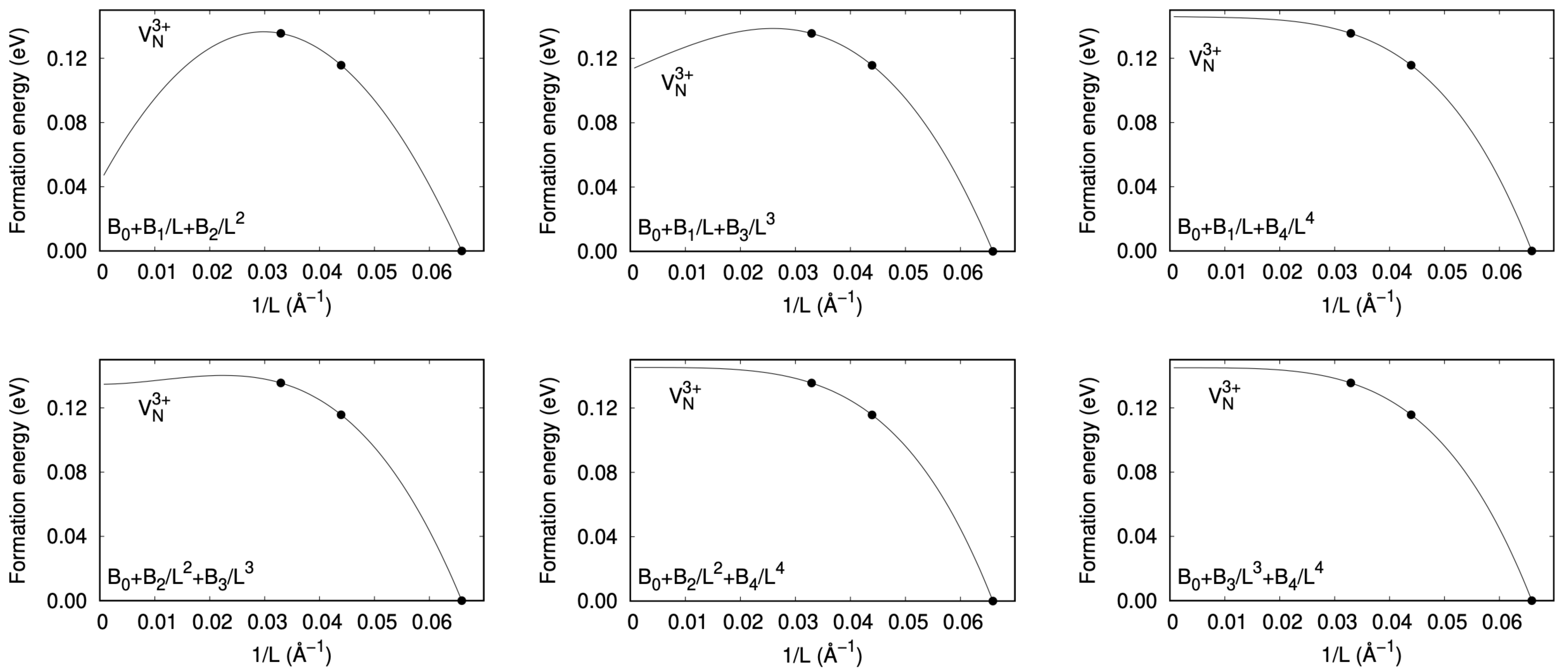}
\caption{Fittings for the formation energy of V$_\textrm{N}^{3+}$ using different polynomials. 
The corrections are, from left to righ, and top to bottom: 0.043~eV, 0.113~eV, 0.146~eV, 0.135~eV, 0.146~eV, and 0.145~eV. The Makov-Payne correction (top-center) gives $B_{1}>0$ in contradiction with the model. Instead, we choose as correction the difference between $8\times 8\times 8$ and $4\times 4\times 4$ supercells, i.e., 0.136~eV. Alternatively, the linear-quartic fit (top right) gives 0.146~eV.}
\end{figure*}

\clearpage

\section{Charge neutrality equation}

The charge-neutrality equation (CNE) solved to determine the Fermi level ($E_F$) as a function of temperature $T$ ($E_F=E_F (T)$) is given by~\cite{Freysoldt2014SM}
\begin{equation}
\label{ec:neutrality}
p(E_F) - n(E_F) + \sum_{X,q} q N_X e^{-H(X^q, E_F)/k_B T} = 0 ,
\end{equation}
where $H(X^q, E_F)$ is the formation energy of defect $X$ with charge $q$, $N_X=n_X/V$ represents the defect density, $V$ is the volume of the primitive cell, $k_B$ is the Boltzmann constant, and $n_X$ is the number of available sites for the defect within the primitive cell. In this case: $n_{\mathrm{V}_\mathrm{N}} = n_{\mathrm{Cu}_\mathrm{i}} = 1$, $n_{\mathrm{V}_\mathrm{Cu}} = 3$, and $n_{\mathrm{N}_\mathrm{i}} = 6$. The latter corresponds to the $N_i$ locations near each of the six faces of the unit cell.

The electron $n(E_F)$ and hole $p(E_F)$ densities were obtained by numerical integration of the density of states near the conduction band minimum (CBM) and the valence band maximum (VBM) multiplied by the electron and hole occupation function, respectively, i.e.:
\begin{equation}
\label{ec:nef}
n(E_F) = \int_{E_{\mathrm{CBM}}}^{E_{\mathrm{CBM}} + 11 k_B T} D(E) \frac{1}{1 + \exp\left(\frac{E - E_F}{k_B T}\right)} \, dE,
\end{equation}
and 
\begin{equation}
\label{ec:pef}
p(E_F) = \int_{E_{\mathrm{VBM}} - 11 k_B T}^{E_{\mathrm{VBM}}} D(E) \frac{1}{1 + \exp\left(\frac{E_F - E}{k_B T}\right)} \, dE .
\end{equation}
Integrating over a range of $11 k_B T$ from the band edges in Eqs. \eqref{ec:nef} and \eqref{ec:pef} ensures numerical stability. 

The density of each species is the sum of the densities of all charge states for that specific defect: 
\begin{equation}
\rho_X(T) = \sum_{q} { q N_X e^{-H(X^q, E_F(T))/k_B T} }.
\end{equation}

The density of states, $D(E)$ in Eqs. \eqref{ec:nef} and \eqref{ec:pef} has been computed using the VASP code, for the primitive cell of Cu$_3$N. The first Brillouin zone was sampled by means of a $\Gamma$-centered $32\times 32\times 32$ \textbf{k}-point mesh with the tetrahedron method. The HSE functional and a plane-wave kinetic-energy cutoff of 400~eV were used, as described in the main text. Small energy shifts (different for valence and conduction bands) were applied to the energies in the density of states dataset, to set the VBM and the CBM at the values obtained from the ionization potential and electron affinity (see method B in Fig.~3 of the main text), including a size correction.

\section{Temperature and pressure dependence of formation energies}

The formation energy of the copper vacancy is computed as (see Eq. 15 in the main text)
$$H(\mathrm{V}^q_{\Cu})= E(\Cu_3\N+\mathrm{V}^q_{\Cu})-E(\Cu_3\N)+\Delta E_{\mathrm{corr}} + \mu^0_{\Cu} + \Delta \mu_{\Cu} + q(E_{\mathrm{VBM}}+E_F),$$
where $\Delta E_{corr}$ includes potential alignment and size corrections. The chemical potential term satisfies
$$\Delta \mu_{\Cu} = \frac{1}{3}\bigg(\Delta H(\Cu_3\N)- \Delta \mu_\N\bigg).$$
Therefore, a decrease of $\Delta\mu_\N$ increases both $\Delta \mu_{\Cu}$ and the formation energy. For a monoatomic gas-like dissociated N, the chemical potential depends on temperature and pressure as given by
$$\mu(T,P) = E_\N - k_B T \ln \bigg[ g_0 \frac{k_B T}{P} \bigg(\frac{2 \pi m k_B T}{h^2}\bigg)^{3/2} \bigg],$$
where E$_\N$ is the atomic energy in its ground state and $g_0$ is the degeneracy of the ground state. Therefore,
$$\Delta \mu_\N = - k_B T \ln \bigg[ g_0 \frac{k_B T}{P} \bigg(\frac{ m k_B T}{2 \pi \hbar ^2}\bigg)^{3/2} \bigg] = k_B T \ln \bigg[ \frac{P}{k_B T} \bigg(\frac{2 \pi \hbar ^2}{ m k_B T}\bigg)^{3/2}\bigg] - k_B T \ln g_0.$$

The nitrogen atom in the ground state has three 2$p$ orbitals occupied by three electrons with the same spin projection; therefore, the orbital angular momentum is $L=0$, the spin $S=3/2$, and the total angular momentum is $J=3/2$. Hence, the electronic degeneracy is $g_0 = 2J +1 = 4$.

$\Delta\mu_\N$ is negative, decreases with temperature as $-(5/2)k_B T \ln T$, increases with pressure as $k_B T \ln P$, and tends to zero when the temperature approaches zero. This mechanism allows \defe{V}{Cu} to have a positive formation energy in N-rich conditions. Table~S3 shows the values obtained for growth temperatures up to the Cu$_3$N melting point, for a typical pressure of 2.0~Pa (15~mtorr)~\cite{Chen2022SM, Wen2025SM}.

\begin{table}
\addcontentsline{toc}{table}{Table S3}
\centering
\caption{Relative chemical potentials of nitrogen and copper at a pressure of $2.0$ Pa. The values at zero temperature are $\Delta\mu_{\mathrm{N}} = 0$ and $\Delta\mu_{\mathrm{Cu}} = -1.37$ eV.}
\begin{tabular}{c c c c c c c c c c c}
\hline
{$T$ (K)}
& 300 & 330 & 360 & 390 & 420 & 450 & 480 & 510 & 540 & 570 \\
\hline
{$\Delta\mu_{\mathrm{Cu}}$ (eV)}
& -1.14 & -1.11 & -1.09 & -1.06 & -1.04 & -1.01 & -0.98 & -0.96 & -0.93 & -0.91 \\
{$\Delta\mu_{\mathrm{N}}$ (eV)}
& -0.69 & -0.77 & -0.84 & -0.92 & -1.00 & -1.08 & -1.16 & -1.23 & -1.31 & -1.39 \\
\hline
\end{tabular}
\end{table}



\end{document}